\documentclass[12pt,a4paper]{article}
\pdfoutput=1
\usepackage{color}
\usepackage{amssymb,amsmath,bm,bbold,slashed}
\usepackage{epsf}
\usepackage{epsfig}
\usepackage{relsize}
\usepackage[dvipsnames]{xcolor}
\usepackage[linktoc=page,bookmarks=false,colorlinks=false,linkbordercolor=RoyalBlue,citebordercolor=ForestGreen,urlbordercolor=CornflowerBlue]{hyperref}
\usepackage{latexsym,mathrsfs,dsfont}
\usepackage[normalem]{ulem} 
\usepackage[compress]{cite}
\usepackage{graphicx}
\usepackage{url}
\usepackage{booktabs}
\usepackage{float}
\usepackage{multirow}
\usepackage{changepage}
\usepackage[hypcap]{caption, subcaption}
\usepackage{bbold}

\usepackage[T1]{fontenc}
\usepackage[utf8]{inputenc}

\usepackage[titles]{tocloft}

\usepackage{tabularx,colortbl}

\makeatletter
\newcommand{\overleftrightsmallarrow}{\mathpalette{\overarrowsmall@\leftrightarrowfill@}}
\newcommand{\overrightsmallarrow}{\mathpalette{\overarrowsmall@\rightarrowfill@}}
\newcommand{\overleftsmallarrow}{\mathpalette{\overarrowsmall@\leftarrowfill@}}
\newcommand{\overarrowsmall@}[3]{%
  \vbox{%
    \ialign{%
      ##\crcr
      #1{\smaller@style{#2}}\crcr
      \noalign{\nointerlineskip}%
      $\m@th\hfil#2#3\hfil$\crcr
    }%
  }%
}
\def\smaller@style#1{%
  \ifx#1\displaystyle\scriptstyle\else
    \ifx#1\textstyle\scriptstyle\else
      \scriptscriptstyle
    \fi
  \fi
}
\makeatother
\newcommand{\leftrightvec}[1]{\overleftrightsmallarrow{#1}}

\usepackage{fancyhdr}
\advance \headheight by 3.0truept       

\allowdisplaybreaks[1]

\definecolor{red}{cmyk}{0,1,1,0.4}
\definecolor{darkgreen}{rgb}{0.0,0.6,0.0}
\definecolor{cDarkGrey}{RGB}{91,91,91}
\definecolor{cGrey}{RGB}{245,243,238}
\definecolor{cBlue}{RGB}{0,110,191}
\definecolor{cLightBlue}{RGB}{214,237,252}
\definecolor{cRed}{RGB}{196,0,100}
\definecolor{cLightRed}{RGB}{254,222,237}
\definecolor{cGreen}{RGB}{0,166,80}
\definecolor{cLightGreen}{RGB}{254,222,237}
\definecolor{cOrange}{RGB}{221,74,44}
\definecolor{cLightOrange}{RGB}{255,215,210}
\definecolor{cPurple}{RGB}{93,35,125}
\definecolor{cLightPurple}{RGB}{241,230,252}
\definecolor{cYellow}{RGB}{252,191,10}
\definecolor{cISSRBlue}{RGB}{0,111,174}
\definecolor{cISSRGrey}{RGB}{167,169,172}

\newcommand{\bea}{\begin{eqnarray}}
\newcommand{\eea}{\end{eqnarray}}

\newcommand{\GeV}{\,\text{GeV}}

\definecolor{cBlue}{RGB}{0,110,191}
\definecolor{cLightBlue}{RGB}{214,237,252}
\definecolor{cRed}{RGB}{196,0,100}
\definecolor{cLightRed}{RGB}{254,222,237}
\definecolor{cGreen}{RGB}{0,166,80}

\makeatletter
\renewcommand\paragraph{\@startsection{paragraph}{4}{\z@}%
  {-3.25ex\@plus -1ex \@minus -.2ex}%
  {1.5ex \@plus .2ex}%
  {\normalfont\normalsize\bfseries}}
\makeatother

\cftsetindents{section}{0em}{2em}
\cftsetindents{subsection}{2em}{3em}
\cftsetindents{subsubsection}{5em}{4em}

\allowdisplaybreaks[1]

\begin{document}

\vspace{-1cm}

\begin{center}
{\large\bf
\boldmath{Standard Model prediction of the $B_c$ lifetime}}
\\[1.2cm]
{\bf
  Jason~Aebischer and
  Benjam\'in Grinstein
}\\[0.5cm]

{\small
Department of Physics, University of California at San Diego,
    La Jolla, CA 92093, USA
}
\\[0.5 cm]
\footnotesize
E-Mail:
\texttt{jaebischer@physics.ucsd.edu},
\texttt{bgrinstein@ucsd.edu}
\\[0.2 cm]
\end{center}

\vskip 1.0cm

\begin{abstract}
\noindent Applying an operator product expansion approach we update the Standard Model prediction of the $B_c$ lifetime from over 20 years ago. The non-perturbative velocity expansion is carried out up to third order in the relative velocity of the heavy quarks. The scheme dependence is studied using three different mass schemes for the $\bar b$ and $c$ quarks, resulting in three different values consistent with each other and with experiment. Special focus has been laid on renormalon cancellation in the computation. Uncertainties resulting from scale dependence, neglecting the strange quark mass, non-perturbative matrix elements and parametric uncertainties are discussed in detail. The resulting uncertainties are still rather large compared to the experimental ones, and therefore do not allow for clear-cut conclusions concerning New Physics effects in the $B_c$ decay.
\end{abstract}

\setcounter{page}{0}
\thispagestyle{empty}
\newpage

\setcounter{tocdepth}{2}
\setlength{\cftbeforesecskip}{0.21cm}
\tableofcontents

\newpage

%
%
%

\section{Introduction}
The $B_c$ meson is the lightest state with both ``naked'' beauty and
charm.  As such it is stable against both strong and electromagnetic
decay. Its weak decay can proceed through three distinct mechanisms:
either $\bar b$-quark decay, $c$-quark decay or $\bar b$-$c$
annihilation. Experimentally, the $B_c$ lifetime has been measured by LHCb~\cite{Aaij:2014bva,Aaij:2014gka} and CMS~\cite{Sirunyan:2017nbv}
with a world average of \cite{Agashe:2014kda}:
\begin{equation}
  \label{eq:tauExp}
\tau_{B_c}^{\rm exp}=0.510(9)~\text{ps}\,,
\end{equation}
which corresponds to a decay width
\begin{equation}
  \label{eq:GExp}
\Gamma_{B_c}^{\rm exp}=1.961(35)~\text{ps}^{-1}\,.
\end{equation}

Since both valence quarks in the $B_c$ meson are heavy, the state is
similar in structure to it's quarkonium cousins, the $\eta_c$ and
$\eta_b$ pseudoscalar mesons, the lightest members of the $J/\psi$ and
$\Upsilon$ towers of states. This circumstance allows for an effective
treatment in terms of Non-Relativistic QCD (NRQCD), where (much as in
the case of heavy quarkonium) the anti-quark corresponding to the
valence quark, or the quark corresponding to the valence anti-quark,
is integrated out at the respective scale. This method has been used
to estimate the lifetime of the $B_c$ meson in the Standard Model
(SM)~\cite{Bigi:1995fs,Beneke:1996xe,Chang:2000ac}.

The most precise value has been obtained in~\cite{Beneke:1996xe} to
be $\tau_{B_c}=0.52\,\text{ps}$, with an uncertainty from varying the
input charm quark\footnote{More precisely, Ref.~\cite{Beneke:1996xe}
  varies the charm mass in the range
  $1.4\,\text{GeV}<m_c<1.6\,\text{GeV}$, and then fixes the b-quark mass
  by the requirement that the $B_d$ lifetime is reproduced.} mass that results in
$0.4\,\text{ps}<\tau_{B_c}<0.7\,\text{ps}$, corresponding to
\begin{equation}\label{eq:tauBc}
\Gamma_{B_c}=1.92^{+0.58}_{-0.49}~\text{ps}^{-1}\,,
\end{equation}
exclusive of other sources of uncertainty that include, among others, $\pm22\%$
from scale uncertainty in the perturbative calculation.

Similar results were found in other OPE
calculations~\cite{Bigi:1995fs,Chang:2000ac} as well as using QCD sum
rules~\cite{Kiselev:2000pp} or potential
models~\cite{Gershtein:1994jw}. A comparison of the different predictions can be found in~\cite{Gouz:2002kk}.

The experimental measurement in Eq.~\eqref{eq:tauExp} has a much smaller uncertainty than the theory prediction in
Eq.~(\ref{eq:tauBc}). This motivates a reinvestigation of the SM
prediction from over 20 years ago with the goal of improving the
theoretical precision, and the eventual hope of a
precision comparable to the experimental one.

Renewed interest in the $B_c$ lifetime has arisen because it is
susceptible to New Physics (NP) effects. Consequently a more precise
SM prediction allows to place stronger constrains on NP
models. In particular, experimentally measured deviations from SM
expectations in the semileptonic decays $B\to D\tau\nu$,
$B\to D^*\tau \nu$ and $B_c\to \psi\tau\nu$ suggest NP contributions
to the quark level process $b\to c\tau\nu$
\cite{Lees:2012xj,Lees:2013uzd,Huschle:2015rga,Sato:2016svk,Aaij:2015yra,Hirose:2016wfn,Aaij:2017tyk}. These
so-called $R(D)$, $R(D^*)$ and $R(B_c)$ anomalies can be accounted for by
several extensions of the SM. If the NP is realized as an effective
pseudoscalar interaction, that is, a four-fermion interaction involving a
pseudoscalar hadronic bilinear times a leptonic $S-P$ bilinear, then
the $B_c$ lifetime is especially effective in placing constraints on
its strength \cite{Li:2016vvp,Alonso:2016oyd}.  This type of
interaction is often found in models of NP proposed in the
interpretation of the $R(D^{(*)})$ anomalies, such as the two-Higgs-doublet model (2HDM) and
leptoquarks. The $B_c$ lifetime constraint rules out any of the 2HDM
interpretations of $R(D^*)$, including the type III versions of the
model that contain general Yukawa couplings to the different fermions
\cite{Branco:2011iw}, which can explain simultaneously $R(D)$ and
$R(D^*)$,
\cite{Crivellin:2012ye,Crivellin:2013wna,Cline:2015lqp,Kim:2015zla,Crivellin:2015hha,Wang:2016ggf},
going beyond the more restricted set of models
\cite{Hou:1992sy,Tanaka:1994ay,Kiers:1997zt,Chen:2006nua} on which
BaBar \cite{Lees:2012xj, Lees:2013uzd} and Belle \cite{Sato:2016svk}
place constraints.
Leptoquark models have also been proposed to explain these anomalies
and they are susceptible to these bounds when they generate sizable
pseudo-scalar operators \cite{Alonso:2015sja,Li:2016vvp}.

In this work we follow the OPE methodology of
Ref.~\cite{Beneke:1996xe} (henceforth ``BB''). There are several ways
in which we can improve the result of that work. First and foremost,
we use and compare three mass
  schemes to eliminate pole (or ``on-shell'') masses in favor of well
  defined masses and therefore eliminate renormalon ambiguities that
  arise in the on-shell scheme. The largest source of uncertainty in
the calculation of the width is from the pole masses, since the width
scales as the fifth power of these.  BB use {\it ad hoc} values for
$b$ and $c$ pole masses. We perform the calculation in the
  $\overline{\text{MS}}$ mass scheme, the ``Upsilon'' scheme
  \cite{Hoang:1998ng,Hoang:1998hm}, and a ``meson'' scheme that
  incorporates some aspects of the Upsilon scheme. We will expand on
  these below but for now we point out that the Upsilon scheme organizes the
  perturbative expansion in a manner that not only eliminates
  renormalon ambiguities from the expression for the decay width but
  in addition is empirically seen to have better convergence than
  other schemes. Moreover, the masses of the $\Upsilon$ and $J/\psi$
particles, used as inputs, are determined accurately. Second, for the
one-loop calculation of the subprocess $b\to c\bar c s$ BB uses the
result of Bagan {\it et al} \cite{Bagan:1994zd,Bagan:1994qw}. Later,
Krinner, Lenz and Rauh (henceforth KLR) inferred the presence of typos in
the analytic expressions of Bagan {\it et al} \cite{Krinner:2013cja},
because those expressions do not produce the numerical results and
graphs they present. KLR compute the 1-loop corrections anew and agree
with the numerical results of Bagan {\it et al}, inferring thus the
presence of typos. We have used, and verified partially, the results
of KLR.  Third, BB neglect, and we include, the contribution of
penguin operators, that are formally of the same order as other
contributions retained in the calculation albeit numerically
small. Fourth, we use better determined input parameters, such
as the strong coupling constant $\alpha_s(M_Z)$, the CKM matrix
elements, and the non-perturbative $B_c$ decay constant, $f_{B_c}$, for
which there is now a lattice calculation \cite{McNeile:2010ji},
among others.
Fifth, we use spin symmetry to relate some of  the non-perturbative matrix elements appearing in the calculation.

In addition to updating and improving on the BB result, we have tried to  clarify several
issues in their presentation. Among these we clarify
below the need to go up to  fourth order  in velocity in the
NREFT, the role of spin symmetry in the
calculation of matrix elements, the relative (un)importance of
various corrections, the distinction between quarks and anti-quarks in
the NREFT as well as the interpretation and precise value of the momenta $p_b\pm
p_c$ that enter the operators in the OPE for weak annihilation (WA)
and Pauli interference (PI) diagrams.

The rest of the paper is organized as follows: In
Sec.~\ref{sec:Mscheme} the different mass schemes used for the
calculation are discussed. In Sec.~\ref{sec:Heff} we outline the
Effective Field Theory approach used at the electroweak (EW) scale
involving the effective Hamiltonian. The matching onto NRQCD is
performed in Sec.~\ref{sec:NRQCD}. The Operator Product Expansion
(OPE) is discussed in Sec.~\ref{sec:OPE}. Section~\ref{sec:MEs}
describes the computation of relevant matrix elements and in
Sec.~\ref{sec:num} the numerical computation and analysis of the uncertainties for the SM prediction is
performed before we summarize our results and offer some  observations in the
conclusions,  in Sec.~\ref{sec:cocl}.

\section{Mass schemes}
\label{sec:Mscheme}
The largest source of uncertainty in the calculation of the inclusive
rate of the $B_c$ decay is in the value of the pole mass. As we will
review below, the calculation of the rate, $\Gamma_{B_c}$, is based on an OPE of the two point function of the effective
Hamiltonian, cf Eqs.~\eqref{eq:GammaTran}
and~\eqref{eq:TranOpDef}. The leading term in the calculation corresponds to $\Gamma_b+\Gamma_c$, the sum of
decay rates of $b$ and $c$ quarks computed perturbatively as if the quarks were not bound to the
$B_c$ meson. The rate $\Gamma_q$ is given in terms of the pole
mass, $m_q$,  as an expansion in powers of $\alpha_s$
\begin{equation}
  \label{eq:QratePole}
  \Gamma_{q}=\kappa m_q^5( f_0(x)+\epsilon\alpha_s f_1(x)+\epsilon^2\alpha_s^2 f_2(x)+\cdots)\,,
\end{equation}
where $\kappa$ is a constant
  independent of $m_q$, $x$  and $\alpha_s$ chosen so that the
  tree-level rate has $f_0(0)=1$.
For reasons that will be explained momentarily, we have introduced the power counting parameter
$\epsilon=1$ that is set to unity at the end of the calculation. The coefficients of the
expansion, $f_i(x)$ are functions of the ratios of final state pole masses, $m_{q'}$, to the mass of the
decaying quark, $x=m_{q'}/m_q$. Pole masses are convenient for the perturbative calculations, but are beset by both computational
and conceptual difficulties: their perturbative expansion is poorly convergent and suffers from a
renormalon ambiguity. Moreover, the roots of the implicit equation that relates them to short distance
(Lagrangian) masses are complex for the light quarks.

 When expressed in terms of the pole mass, the perturbative
 rate also suffers from a renormalon ambiguity.  Remarkably, eliminating the pole mass in favour of
 well defined  ({\it e.g.}, short distance) masses, gives a perturbative expansion of
 the rate that  is free of renormalon
 ambiguities \cite{Beneke:1994bc,Luke:1994xd,Neubert:1994wq,Sinkovics:1998mi}. For each choice of well
 defined mass the manner in which renormalon ambiguities cancel in the rate suggests how to
 organize the perturbative expansion. We refer to any one choice of mass and of reorganization of
 the perturbative series as a ``mass-scheme''. When the well defined
 quark masses are chosen as the modified minimally
 subtracted masses, $\overline{m}_q$, the expansion of the pole mass takes the form
 \begin{equation}
    \label{eq:mpoleMSExp}
   m_q=\overline{m}_q(1+\epsilon \alpha_s d_q^{(1)} +\epsilon^2 \alpha_s^2 d_q^{(2)} +\cdots)\,,
 \end{equation}
 so that the renormalon free expansion for the rate for the flavour transition $q\to q'$ takes the form
 \begin{equation}
 \Gamma_{q\to q'}= \kappa   \overline{m}_q^5\left[f_0(\bar x) + \epsilon\alpha_s\left(5
     d_q^{(1)}f_0(\bar x) +( d_{q'}^{(1)} - d_q^{(1)} ) \bar x \frac{df_0(\bar x)}{d\bar x}+f_1(\bar x)  \right)+\mathcal{O}(\epsilon^2) \right]\,,
\end{equation}
where $\bar x=\overline m_{q'}/\overline m_q$. As we see, in this
``$\overline{\text{MS}}$ scheme'' the expansion in $\epsilon$ is
equivalent to the perturbative expansion in powers of $\alpha_s$.

By contrast, in the ``Upsilon scheme'' the power counting in
$\epsilon$ does not correspond to powers of $\alpha_s$.  Up to small
non-perturbative corrections the $q$ pole mass is given in terms of
that of the mass $M_{\bar q q}$ of some  quarkonium $\bar q q$ state by
\begin{equation}
  \label{eq:mpoleUpsExp}
  m_q=\tfrac12M_{\bar q q}(1+\epsilon \alpha_s^2 D_q^{(1)} +\epsilon^2 \alpha_s^3 D_q^{(2)} +\cdots)\,.
  \end{equation}
For example, if in $\Gamma_{b\to c ud}$ we neglect the light quark masses and use this scheme for
both $b$ and $c$ quarks, then
\begin{multline}
  \label{eq:bRateUpsExp}
 \Gamma_{b\to cud}= \kappa   \left(\frac{M_{\bar b b}}2\right)^5\left\{f_0(X) + \epsilon\left[\alpha_s^2\left(5
     D_b^{(1)}f_0(X) +( D_c^{(1)} - D_b^{(1)} ) X \frac{df_0(X)}{dX}\right)+\alpha_sf_1(X)
 \right]\right.\\
\left. \phantom{\frac{df_0(X)}{dX}}+\mathcal{O}(\epsilon^2) \right\}\,,
\end{multline}
where now $X=M_{\bar c c}/M_{\bar b b}$.

 While any good choice of well defined masses
yields a well behaved perturbative expansion in the sense that it is
free of renormalon ambiguities, different mass-schemes may differ in how
rapidly the expansion converges (in the asymptotic expansion
sense). Were we able to compute to high order,  all
mass-schemes would give the same  numerical value for the rate (up to
a small higher order term).  But  computations of rates are available only to low orders in
perturbation theory, often only including 1-loop corrections to the
leading, tree level term.  In
practical term it is best to choose a scheme that converges most rapidly. In the sub-sections below
we present and compare the three
schemes in which we perform calculations.

As stated above, the leading term in the  OPE for the decay width of $B_c$ mesons is the
perturbative  $\Gamma_b+\Gamma_c$. The corrections to those are expressed as products of non-perturbative matrix
elements and Wilson coefficients. The latter are perturbatively computed as functions
of pole masses. In the calculations below we use the same scheme choice for these sub-leading
terms as for the leading ones.

\subsection{The $\overline{\text{MS}}$ mass-scheme}
This scheme uses the running renormalized Lagrangian masses $\overline m_b(\mu)$ and
$\overline m_c(\mu)$  evaluated at a sufficiently high renormalization scale $\mu$. In this scheme the expansion parameter
$\epsilon$ simply counts powers of $\alpha_s$, making its use particularly
simple. The expansion of $m_{b,c}$  in terms of the
$\overline{\text{MS}}$ masses is known to third order in $\alpha_s$
\cite{Gray:1990yh , Marquard:2007uj}, but  we only need it to first order:
\begin{equation}
  \label{eq:poleM1loop}
  m_q=\overline{m}_q(\mu)\left[1+\epsilon\frac{\alpha_s(\mu)}{\pi}\left(\frac{4}{3}-\ln\left(\frac{\overline{m}_q(\mu)^2}{\mu^2}\right)\right)\right]+\mathcal{O}(\alpha_s^2)\,,
\end{equation}
where we have retained explicitly  the power of $\epsilon$ used in organizing the perturbative
expansion, corresponding to Eq.~\eqref{eq:mpoleMSExp}. It is convenient to cast the $c$-quark decay rate in terms of the running mass evaluated at
itself, $\overline{m}_c(\overline{m}_c)$, and the $b$-quark decay rate in terms of both quark
masses evaluated at the shorter distance scale,
$\overline{m}_b(\overline{m}_b)$ and $\overline{m}_c(\overline{m}_b)$. The value of the masses, $\overline{m}_b(\overline{m}_b)$ and
$\overline{m}_c(\overline{m}_c)$, is reported by the PDG with 1\%-2\%
accuracy. Because the rates scale as the fifth power of the mass, this
results in an uncertainty   of 5\%-10\%. In addition, as discussed below,  the  convergence of
the perturbative series is slow in the case of
semileptonic $b$ and $c$ decays, and there is no reason to suspect it is any better for non-leptonic decays.

\subsection{The Upsilon scheme}
The Upsilon expansion was introduced by Hoang, Ligeti and Manohar in
Refs.~\cite{Hoang:1998ng,Hoang:1998hm} (henceforth ``HLM'') to address the largest source
of uncertainty in the calculation of the inclusive rate of semileptonic $B$ and
$D$ meson decays. The calculation of the rate for semileptonic $B$
decays  is based on an operator
product expansion of the two-point function of hadronic charged
currents
\[
i\int d^4x\, e^{iq\cdot x}  \langle B|  T( j^\mu(x)
j^{\nu\dagger}(0)) |B\rangle
\]
in terms of operators in the Heavy Quark Effective Theory (HQET)
\cite{Chay:1990da}. The expansion is naturally given in inverse powers
of the pole mass of the decaying heavy quark, and  the resulting
rate is proportional to the fifth power of the pole mass. In the
Upsilon scheme the masses are chosen to be the well measured 1S
masses,\footnote{The masses of other quarkonium states can be used
  instead.}  $M_{\bar bb}=m_\Upsilon$ and $M_{\bar cc}=m_{J/\psi}$, resulting in a
negligible uncertainty in the decay rate  from the value of these masses. The expansion
of the pole masses in terms of $m_\Upsilon$ and $m_{J/\psi}$ has been determined to
fourth order in $\alpha_s$, with corrections that start at order
$\alpha_s^2$ \cite{Pineda:1997hz,Melnikov:1998ug}:
\begin{equation}
  \frac{\tfrac12m_{\Upsilon}}{m_b}=1-\frac{(\alpha_s C_F)^2}{8}
  \left\{1\epsilon
    +\frac{\alpha_s}{\pi}\left[\left(\ln\left(\frac{\mu}{\alpha_sC_Fm_b}\right)+\frac{11}{6}\right)\beta_0-4\right]\epsilon^2+\cdots\right\}\,,
\end{equation}
where $\beta_0=11-\frac23 n_f$  is the
coefficient of the first term in the QCD  $\beta$-function,
$C_F=(N_c^2-1)/2N_c=4/3$, and we have omitted the known term of order
$\alpha_s^4$. For quarkonium systems the quark potential suffers from a
renormalon ambiguity, as does the pole mass, but the resulting quarkonium
state energy is free from ambiguities and hence the onium mass is a
good candidate for an unambiguous, well defined mass \cite{Hoang:1998nz,Beneke:1998rk}.

The organization of the expansion is unusual.  The parameter
$\epsilon=1$ is inserted to indicate the order in the
expansion. For the expansion of the rate in Eq.~\eqref{eq:QratePole} the
power of $\epsilon$ matches that of $\alpha_s$.  But, as seen
above, the parameter $\epsilon$ in Eq.~\eqref{eq:mpoleUpsExp} carries
one additional power of $\alpha_s$, {\it i.e.},  $\epsilon^n\alpha_s^{n+1}$. As seen in Eq.~\eqref{eq:bRateUpsExp}
the leading correction to the rate includes order $\alpha_s$ terms
from the perturbative corrections to the rate, and order $\alpha_s^2$
from the perturbative expansion of the masses. This is dictated by the
requirement that the renormalons cancel in the expression for the
rate.

In the implementation of this scheme for non-leptonic decays HLM
expand the Wilson coefficients of the electroweak effective
Hamiltonian in powers of $\alpha_s\ln(m_W/m_b)$ and truncate in the
power of $\alpha_s$ to which they perform the calculation of the
rate. We do not adopt this prescription and instead retain the full
resummed values of the Wilson coefficients. The reason is that this
series does not contribute to the cancellation of renormalons, as can
be seen from the fact that $M_W$ is an arbitrary parameter. This is
the case for the other mass schemes as well.

\subsection{The ``meson'' scheme}
The HQET gives the heavy baryon masses  as an expansion in inverse
powers of the pole masses.  To second order in the heavy quark expansion one has
\begin{equation}
  \label{eq:poleMassDiff}
  m_b-m_c=\overline{m}_B-\overline{m}_D+\frac12\lambda_1\left(\frac1{m_b}-\frac1{m_c}\right)
  +\cdots
\end{equation}
This is written in terms of the spin- and isospin-averaged masses
$\overline{m}_B=\frac14(3 m_{B^*}+m_B)$ and
$\overline{m}_D=\frac14(3 m_{D^*}+m_D)$, and the
non-perturbative HQET parameter
$\lambda_1=$ \break $-\frac12\langle H^Q|\bar
Q_v D_\perp^2 Q_v|H^Q\rangle$, where $Q_v$ is a heavy quark with
4-velocity $v$ and $H^Q$ stands for the heavy-light meson in the HQET;
for details see, {\it e.g.} Ref.~\cite{Manohar:2000dt}. Using the spin
average mass eliminates dependence on the second non-perturbative
parameter that enters at this order, $\lambda_2$,  that is responsible for the mass splitting of
the $B-B^*$ and $D-D^*$ heavy quark spin multiplets. In the meson scheme one eliminates the charm mass
$m_c$ in favour of $m_b$, the very well measured physical meson masses and
non-perturbative parameters of the heavy quark expansion. For $m_b$
one may then choose one of the previous schemes; for our work we  take $m_b$ from $m_\Upsilon$ as in
the Upsilon scheme above.

Notably, the leading correction to the relation between pole and meson masses, given by the non-perturbative
parameter $\bar\Lambda$, cancels out in the mass difference in
Eq.~\eqref{eq:poleMassDiff}, improving the precision and eliminating
the renormalon in $\bar\Lambda$. Another important advantage of this
scheme is that the ``phase space'' functions $f_i(x)$ incorporate the
dependence on the mass difference which is now fixed accurately.

If one chooses, as we do, to give $m_b$ in terms of  $m_\Upsilon$ then the organization of the
expansion in the meson scheme is analogous to that of the Upsilon scheme. Instead of Eq.\eqref{eq:mpoleUpsExp}  we now have
\begin{align}
  m_c&=m_c^{(0)}+\epsilon\alpha_s^2m_c^{(1)}+\cdots \nonumber
  \\
  &= \frac{m_\Upsilon}{2}-\overline{m}_B+\overline{m}_D
  -\frac12\lambda_1\left(\frac2{m_\Upsilon}-\frac2{m_\Upsilon -2(\overline{m}_B-\overline{m}_D)}\right)\nonumber
  \\
&\qquad\qquad\qquad + \epsilon\alpha_s^2 C_b^{(1)}\frac{m_\Upsilon}{2}\left[ 1+2\lambda_1\left(\frac1{m_\Upsilon^2}-\frac1{(m_\Upsilon -2(\overline{m}_B-\overline{m}_D))^2}\right)\right] +\cdots
  \end{align}
and the rate expansion takes a form similar to that of Eq.~\eqref{eq:bRateUpsExp}
\begin{multline}
  \label{eq:bRateMesExp}
  \Gamma_{b\to cud}= \kappa   \left(\frac{m_\Upsilon}{2}\right)^5\bigg\{f_0(X) \\
 + \epsilon\left[\alpha_s^2\left(5
     C_b^{(1)}f_0(X) +2\left( \frac{m_c^{(1)}}{m_\Upsilon} - C_b^{(1)}\frac{m_c^{(0)}}{m_\Upsilon} \right)  \frac{df_0(X)}{dX}\right)+\alpha_sf_1(X)
 \right]\\
+\mathcal{O}(\epsilon^2) \bigg\}\,.
\end{multline}

\subsection{Light quarks}\label{sec:lightquarks}
The available 1-loop calculations give  decay rates in terms of pole masses. At $L$-loops the pole
mass $m_q$ is determined implicitly, as a root of the equation
\begin{equation}
  \label{eq:mpole}
\frac{m_q}{\overline{m}_q(m_q)}=1+\sum_{n=1}^L d_n\left(\frac{\alpha_s(m_q)}{4\pi}\right)^n\,,
\end{equation}
where the $\overline{\text{MS}}$-mass, $\overline m_q(\mu)$, and the running coupling,
$\alpha_s(\mu),$ are computed at $L$-loops~\cite{Gray:1990yh}.\footnote{A compendium of useful
  formulae for on-shell masses can be found in Ref.~\cite{Chetyrkin:2000yt}.}  Beyond 1-loop this relation has only
complex roots for small input $\overline{\text{MS}}$-mass; that is,  the relation
involves necessarily $\ln(\ln(m_q/\Lambda_{\text{QCD}}))$ which is complex for the light quarks,
$q=u,d,s$. The effect of including the light quark masses  is expected to be small, so one may get
around this difficulty by working in the massless limit; chiral symmetry guarantees, perturbatively,
the vanishing of the pole mass. Yet, the strange quark is sufficiently heavy as to have a
significant effect on charm decays. Therefore, in principle Eq.~\eqref{eq:mpole} has to be solved iteratively in order to
give $m_q$ in terms of $\overline m_q(\mu)$ at a sufficiently high scale $\mu$, where perturbation
theory is still valid. In this work however we only use the 1-loop result given in Eq.~\eqref{eq:poleM1loop} together with the results from Monte Carlo simulations of QCD
on the lattice that reliably  determine $\overline{m}_q(\mu)$ at $\mu=2$~GeV. For the
expansion in $\epsilon$ we use the power counting indicated in Eq.~\eqref{eq:poleM1loop}, regardless
of the scheme used for the heavy quarks.

The numerical impact of the non-zero strange quark mass will be discussed below. We use a non-zero
strange mass for charm decays and neglect $m_s$ in $\bar b$-decays as well as in WA and
PI. Clearly the non-vanishing mass tends to
decrease rates as it restricts, if only marginally, the phase space available to the decay
products. But this effect is very small in decays of the heavier $b$ quark and even smaller in WA. Because we do not consistently include a non-zero strange quark  mass in all $B_c$ decay channels, we give
our results using a vanishing strange quark mass in all $B_c$ decay channels, and then, in addition, we list
the results of decay channels from charm decay anew with non-zero strange quark mass.

\subsection{Nomenclature}
For clarity we summarize our definitions of ``schemes'' as applied to the computation of the $B_c$
lifetime in the rest of this work:
\begin{description}
\item[$\overline{\text{MS}}$] For all partial rates we use $m_b$ and $m_c$ given in terms of $\overline
  m_b(\mu_b)$ and $\overline
  m_c(\mu_c)$ respectively.
\item[Meson] For all partial rates we use
    Eq.~\eqref{eq:poleMassDiff} to give $m_c$    in terms of $m_b$
  and $\overline{m}_{B}-\overline{m}_{D}$, and we use the Upsilon expansion  to give $m_b$ in terms of
  $m_\Upsilon$.
\item[Upsilon] The contributions to the total width that arise from $\bar b$ decays, WA
  and PI are computed as in the meson scheme, but for those from $c$ decays
  the $c$-quark pole mass is given by the Upsilon expansion in terms of~$m_{J/\psi}$.
\end{description}
The contributions to the total width arising from $\bar b$ decays, WA
  and PI are computed with $\mu_b=\mu_c=\overline
  m_b(\overline m_b)$, while for $c$ decays we use $\mu_c=\overline
  m_c(\overline m_c)$. These choices are motivated by the typical total energy released in each of these
  processes. In all three schemes the light quarks, $q=u,d,s$  are assumed to be massless, except for
  the strange quark for which we use
  $m_s$ in terms of $\overline m_s(\mu_c)$ when computing partial $B_c$ decay widths from $c$-quark decays.

\subsection{Comparison of schemes}
\label{sec:schemes-compared}
We have already indicated that both the Upsilon  and the meson
schemes have the advantage that the input masses are extremely well
known. This gives them a clear advantage over the $\overline{\text{MS}}$
scheme. However, one should keep in mind that the relation giving the
pole mass in terms of the 1S mass is subject to poorly known
non-perturbative corrections. The authors of HLM estimate the
non-perturbative mass shift $\delta m$  as $\delta m\sim a^3\Lambda_{\rm QCD}^4$, where
$a$ is the Bohr radius of the 1S. For the Upsilon state they estimate
this correction as $\delta m\sim15~\text{MeV}$ for $\Lambda_{\rm
  QCD}=350~\text{MeV}$ ($\delta m\sim60~\text{MeV}$ for $\Lambda_{\rm
  QCD}=500~\text{MeV}$).  The leading contribution to the mass shift
is from the gluon condensate $\langle \alpha_sG^2_{\mu\nu}\rangle$,
and has been calculated in
Refs.~\cite{Leutwyler:1980tn,Voloshin:1979uv}:
\[
  \delta m =1.41 m_b\frac{\pi \langle
    \alpha_sG^2_{\mu\nu}\rangle}{(m_b C_F \alpha_s(a^{-1}))^4}\,.
\]
This is estimated as $\delta m \approx 60~\text{MeV}$ in
Ref.~\cite{Pineda:1997hz}. HLM uses $\delta m\sim 100~\text{MeV}$ as a
conservative estimate of the error in the pole mass relation. This
seems to result from an abundance of caution. We do not include the
condensate correction in our calculations. Were we to include it in the calculations in the Upsilon
scheme the uncertainty in the charm quark mass would render it
useless. Instead we assume it can be
neglected. This can be checked a posteriori, and much like the
implicit assumption that quark-hadron duality can be used for the
implementation of the OPE method we view this as an assumption that can
be validated either by further theoretical progress or by experimental results.

Turning to the nature of the expansions, we present calculations of
semileptonic decay rates of $B$ and $D$ mesons to compare some of the
schemes.  Using $\epsilon=1$ to indicate orders of the expansion,
the rate for $B\to X_u e\nu$ in the Upsilon scheme behaves as\footnote{HLM
  only include the BLM part of the $\epsilon^2$ term. Since the
  publication of HLM the full 2-loop correction has become
  available \cite{Pak:2008cp,Melnikov:2008qs}. For some semileptonic
  decays HLM compare the Upsilon scheme  to the pole mass
  expansions. We always include the more relevant comparison to the $\overline{\text{MS}}$ scheme. }
\[
  1-0.115\epsilon-0.030\epsilon^2+\cdots
\]
while for the $\overline{\text{MS}}$ scheme
\[
  1+0.30\epsilon+0.20\epsilon^2+\cdots
\]
For the numerical estimate  we have used $\alpha_s=0.223$,
corresponding to  the running
coupling evaluated at $\mu=\overline{m}_b(\overline{m}_b)$. The
ellipses stand for higher orders of   $\epsilon$ as well as
non-perturbative corrections. These
estimates seem to indicate more rapid convergence of the Upsilon  scheme
than for the $\overline{\text{MS}}$ scheme. For comparison we also give the
expansion for the rate in terms of the pole mass:
\[
    1-0.17\epsilon-0.11\epsilon^2+\cdots
\]
The expansion in the Upsilon scheme seems to be approaching a limit
quickly, as each term in the expansion is an order of magnitude
smaller than the previous. By contrast the expansion in the
$\overline{\text{MS}}$ scheme displays much slower convergence. One
expects of course that after a sufficient number of terms are included,
both Upsilon and $\overline{\text{MS}}$ scheme expansions will give
equivalent rates.  But in practice only a small number  of terms in
the series can be computed and it is most practical to use the scheme
that converges fastest. We note that the convergence  in terms of the pole
mass is similar to that of the $\overline{\text{MS}}$ scheme.  This should not suggest
to use the pole mass as a scheme: as mentioned above, both the mass and
rate expansion are ambiguous.

For $B\to X_c e \nu$ we find the following expansions
\begin{align*}
    &1-0.10\epsilon-0.03\epsilon^2+\cdots& &\text{(meson/Upsilon scheme)}\\
    &1+0.27\epsilon+0.09\epsilon^2+\cdots& &\text{($\overline{\text{MS}}$  scheme)}\\
    &1-0.20\epsilon-0.20\epsilon^2+\cdots& &\text{(pole mass)}
\end{align*}
and, using $\alpha_s=0.373$ corresponding to evaluating the running
coupling constant at $\mu = \overline{m}_c(\overline{m}_c)$, we find
for $D\to X_{s+d}e\nu$ the expansions
\begin{align*}
    &1-0.13\epsilon+0.02\epsilon^2+\cdots& &\text{(Upsilon scheme)}\\
    &1+0.51\epsilon+0.64\epsilon^2+\cdots& &\text{($\overline{\text{MS}}$  scheme)}\\
    &1-0.29\epsilon-0.30\epsilon^2+\cdots& &\text{(pole mass)}
\end{align*}
In addition, the reader may find in HLM similar estimates for $B\to
X_{u,c}\tau\nu$, and $B\to X_{c(\bar s+\bar d)}$ (that only retain the BLM
part of the $\epsilon^2$ term). In all cases it is apparent  that the
expansion in the
Upsilon scheme converges faster than in the $\overline{\text{MS}}$
scheme and faster than for the (a priori ill defined) rate in terms of pole masses.

\section{Effective Hamiltonian}\label{sec:Heff}
We employ the standard Effective Field Theory approach where the heavy
SM particles (top quark, Higgs and EW gauge bosons) are integrated out
at the EW scale and matched onto the following effective Hamiltonian
\cite{Buras:1998raa,Buras:2020xsm}:\footnote{At one loop-order the
  electroweak penguin operators $Q_{7}-Q_{10}$ are generated as
  well. We neglect their contributions, since they correspond to
  higher order corrections in our expansion.}
\begin{equation}
  \label{eq:Heff}
  \mathcal{H}_{eff}= \frac{4G_F}{\sqrt{2}}V_{cb}^{\phantom{*}}V_{cs}^*\sum_{i=1}^{6} C_i Q_i +\text{h.c.}\,,
\end{equation}
with the current-current operators
\begin{equation}
  Q_1= (\bar s \gamma_\mu P_L b)(\bar c \gamma^\mu P_L c)\,, \qquad
  Q_2= (\bar s^\alpha \gamma_\mu P_L b^\beta)(\bar c^\beta \gamma^\mu P_L c^\alpha)\,,
\end{equation}
and the QCD-penguin operators
\begin{align}
  Q_3&= (\bar s \gamma_\mu P_L b)\sum_q(\bar q \gamma^\mu P_L q)\,,\qquad
   Q_4= (\bar s^\alpha \gamma_\mu P_L b^\beta)\sum_q(\bar q^\beta \gamma^\mu P_L q^\alpha)\,,\\\notag
   Q_5&= (\bar s \gamma_\mu P_L b)\sum_q(\bar q \gamma^\mu P_R q)\,,\qquad
    Q_6= (\bar s^\alpha \gamma_\mu P_L b^\beta)\sum_q(\bar q^\beta \gamma^\mu P_R q^\alpha)\,.\notag
  \end{align}
Here $P_{R,L}= \frac{1}{2}(1\pm\gamma_5)$ and $\alpha,\beta$ denote colour indices. We
have used
$V_{tb}^{\phantom{*}}V_{ts}^* \approx - V_{cb}^{\phantom{*}}V_{cs}^*
$, which holds to high precision. The sums run over the active quarks
at the given scale.

{To compute the relevant partial decays rates at one-loop, one has to combine one-loop matrix elements together with Wilson coefficients resulting from a one-loop matching and two-loop running calculation. In our approach we only work up to this order for the operators $Q_{1,2}$ but not for the QCD penguin operators. This simplification is justified by the fact that the operators $Q_{3,4,5,6}$ have much smaller Wilson coefficients. For convenience we report in Tab.~\ref{tab:WCs} the LO Wilson coefficients of the operators in eq.~\eqref{eq:Heff} at the scales of the $b$- and $c$-quark.}

\begin{table}
\centering
\renewcommand{\arraystretch}{1.3}
\begin{tabular}{|c|c|c|}
\hline
 Wilson coefficient
& $m_b$ = 4.195 GeV
& $m_c$ = 1.2734 GeV
\\
\hline
  $C_1$                   &  $-0.241$&  $-0.420$
\\
\hline
  $C_2$                   & $1.103$&  $1.205$
\\
\hline
  $C_3$                   & $0.011$  &  $0.021$
\\
\hline
  $C_4$                   & $-0.025$ & $-0.043$
\\
\hline
  $C_5$                   & $0.007$  &  $0.012$
\\
\hline
  $C_6$                   &  $-0.030$ &  $-0.060$
\\
\hline
\end{tabular}

\caption{\small
Wilson coefficients at the $b$- and $c$-scale, taking tree-level matching and one-loop RG running effects into account.
}
  \label{tab:WCs}
\end{table}

The semileptonic channels also contribute to the decay rate of the
$B_c$ meson. Integrating out the $W$ at the EW scale leads to the
following effective charged-current operators:
\begin{equation}
  \label{eq:HSL}
  \mathcal{H}_{eff}^{SL} = \frac{4G_F}{\sqrt{2}}V_{cb}\sum_\ell C_\ell(\bar c\gamma_\mu P_L b)(\bar \ell \gamma^\mu P_L \nu_\ell)\,,
\end{equation}
where the sum runs over all lepton flavours. At tree-level in the SM with the chosen normalization the Wilson coefficients are equal to one for all lepton flavours, $C_\ell = 1$. These operators do not run under QCD due to current conservation and have a small running under QED \cite{Aebischer:2017gaw}. Therefore, we only consider the tree-level matching at the electroweak scale and neglect RG effects for these operators.

With a slight abuse of notation, we use the same symbols to denote operators and Wilson coefficients
of the effective Hamiltonian for charm hadronic decay,
\begin{equation}
  \label{eq:Hceff}
  \mathcal{H}_{eff}^{(c)}= \frac{4G_F}{\sqrt{2}}V_{cs}^*V_{ud}^{\phantom{*}}( C_1 Q_1+C_2 Q_2) +\text{h.c.}\,,
\end{equation}
with the current-current operators
\begin{equation}
  Q_1= (\bar s \gamma_\mu P_L d)(\bar u \gamma^\mu P_L c)\,, \qquad
  Q_2= (\bar s^\alpha \gamma_\mu P_L d^\beta)(\bar u^\beta \gamma^\mu P_L c^\alpha)\,.
\end{equation}
Because these operators carry four separate flavours, there is no QCD mixing from Penguin operators. For semileptonic decays we have the analogue of \eqref{eq:HSL},
\begin{equation}
  \label{eq:HcSL}
  \mathcal{H}_{eff}^{(c)SL} = \frac{4G_F}{\sqrt{2}}V_{cs}\sum_\ell C_\ell(\bar c\gamma_\mu P_L s)(\bar \ell \gamma^\mu P_L \nu_\ell) +\text{h.c.}
\end{equation}

As described in Sec.~\ref{sec:OPE} below, the effective Hamiltonians in Eqs.~\eqref{eq:Heff},
\eqref{eq:HSL}, \eqref{eq:Hceff} and~\eqref{eq:HcSL}  are then used together with an OPE to obtain the $B_c$ lifetime.

\section{Non-Relativistic QCD}\label{sec:NRQCD}
The $b$-quark and $c$-antiquark in the $\bar{B}_c$ meson can be well described in NRQCD. There are
several advantages in utilizing this effective theory. First, it organizes the  computation in an
expansion in powers of  the relative velocity $v=|\vec v|$ of the heavy quarks bound in the
meson. Second,  the expansion makes explicit additional approximate ``spin'' symmetries for the $b$
and $c$ quarks separately. These then give relations among matrix elements that hold even
non-perturbatively (in the QCD coupling expansion).  Third, separate conservation of $b$ and $c$
numbers fixes, non-perturbatively,  the values of  the matrix elements of the leading operators in the OPE that determines
the semi-inclusive partial lifetimes $B_c\to X_{c\bar c}$ and $B_c\to X_{b\bar b}$. The corrections
arise from the order in perturbation theory to which the Wilson coefficients are computed,
and from the higher order terms in the OPE. The matrix elements of the latter  can be estimated
with, say, potential models;  to the extent that the expansion in $v$ is well behaved, the
potentially large uncertainty in the sub-leading matrix elements may still translate in a manageable
uncertainty in the total rate. We briefly review elements of NRQCD.

In our treatment the quark NRQCD fields are still given in terms of Dirac spinors, much like is
commonplace in HQET. The 4-velocity of the quarkonium state is $u^\mu$, with $u^2=1$. We take away
the fast oscillation in the near on-shell evolution of the field
of the QCD heavy quark,  $Q(x)$, of mass $m$, and furthermore project out the positive energy components $\Psi_+$ (that
correspond to the particle annihilation operator),
\[
  Q(x)=e^{-im u\cdot x}(\Psi_+(x)+\Psi_-(x))\,,\qquad\text{where}\qquad
\Psi_\pm=\left(\frac{1\pm\slashed u}{2}\right) \Psi_\pm
\,.\]
The equation of motion, $(i\slashed{D}-m)Q=0$ can then be used to solve for the ``small component''
\[
  \Psi_-=\frac1{2m + iu\cdot D}i\slashed{D}_\perp\,\Psi_+\,,
\]
where for any vector $a^\mu$ the spatial component in the quarkonium restframe is
$a_\perp^\mu=a^\mu-(u\cdot a) u^\mu$. Using this in the QCD Lagrangian for the quark gives
\begin{equation}
  \label{eq:Lag-no-exp}
  \mathcal{L}=\overline\Psi_+\left(iu\cdot D+i\slashed{D}_\perp \frac1{2m + iu\cdot D}i\slashed{D}_\perp\right)\Psi_+
  \,.
  \end{equation}
The NRQCD Lagrangian is obtained by expanding the Lagrangian in Eq.~\ref{eq:Lag-no-exp} in powers of $iu\cdot D/2m$ and truncating the
expansion. The characteristic scale of the derivatives on the field $\Psi_+$ is $\frac12mv^2$ for the
energy, $iu\cdot D$, and $m v$ for the spatial momentum, $D_\perp$. Hence, counting powers of $v$
one has $iu\cdot D\sim D_\perp^2/2m\sim v^2$. In addition, the integral
$\int d^3x\,\bar\Psi_+\Psi_+\sim1$ will be concentrated over the volume $\sim (M v)^{-3}$ of a
quarkonium state of mass $M$, and it follows that $\Psi_+\sim v^{3/2}$. Counting rules for all
fields and derivatives follow from these and the additional field equations
\cite{Lepage:1992tx}.\footnote{Alternatively one can restore the speed of light, $c$, and expand in
  inverse powers, $1/c^n$ \cite{Grinstein:1997gv}. This method shows in addition the requirement to
  incorporate a multipole expansion of the gauge fields. }  In particular, for the gauge field in
Coulomb gauge, $g_s u\cdot A\sim v^2$, $ g_sA_\perp\sim v^3$, which determines the velocity-scaling of the chromo-electric
and magnetic fields (in the quarkonium restframe) $g_s\vec E\sim v^3$ and $g_s\vec B\sim
v^4$. Finally, since the typical momentum is the inverse Bohr radius, one has $\alpha_s(M v)\sim
v$. Expanding the Lagrangian and retaining the lowest order ($\sim v^5$) one has
\begin{equation}
  \label{eq:Lagu^5}
   \mathcal{L}_0=\overline\Psi_+\left(iu\cdot D-\frac1{2m} D^2_\perp\right)\Psi_+\,.
\end{equation}

The NRQCD treatment of antiquark fields is analogous. Now one has
\[
  Q(x)=e^{+im u\cdot x}(X_+(x)+X_-(x))=e^{+im u\cdot x}\left(1+\frac1{2m - iu\cdot D}i\slashed{D}_\perp\right)X_-\,,
\]
where $X_-$ is an antiquark creation operator (containing only negative energy components). It
follows that
\begin{align*}
  \mathcal{L}&=\overline X_-\left(-iu\cdot D+i\slashed{D}_\perp \frac1{2m - iu\cdot
      D}i\slashed{D}_\perp\right)X_- \\
 & =\overline X_-\left(-iu\cdot D-\frac1{2m} D^2_\perp+\cdots\right)X_-\,.
\end{align*}

In order to compare these results to calculations that use the non-relativistic 2-component spinor
notation, and to use estimates of matrix elements that use quark potential models, we recast them in
the rest frame, $u=(1,\vec 0)$. In the Dirac basis of $\gamma$-matrices, $\gamma^0=\sigma^3\otimes
\mathbb{1}$, $\vec \gamma=i\sigma^2\otimes\vec\sigma$, the matrices  $(1+\slashed{u})/2$ and  $(1-\slashed{u})/2$
project out the upper and lower two components of the 4-component spinor, which we denote as $\psi_q$
and $\chi_q$, respectively. Then, for the quark we have
\[
  \mathcal{L}_{\psi}=\psi_q^\dagger\left(iD_t-\frac1{2m}(i\vec D)^2+\frac{g_s}{2m}\vec\sigma\cdot\vec
    B\right)\psi_q\,,
\]
and for the antiquark
\[
  \mathcal{L}_{\chi}=\chi_q^\dagger\left(iD_t+\frac1{2m}(i\vec D)^2-\frac{g_s}{2m}\vec\sigma\cdot\vec
    B\right)\chi_q\,.
\]
While $\psi_q$ is only an annihilation operator, $\chi_q$ is only a creation operator. It is
convenient to rewrite the Lagrangian for the antiquark in terms of the annihilation spinor
\[
  \psi_{\overline q}\equiv i\sigma^2(\chi_q\dagger)^T\,.
\]
In terms of this the anti-quark Lagrangian is
\[
  \mathcal{L}_{\chi}=\psi_{\overline q}^\dagger\left(iD_t-\frac{1}{2m}(i\vec D)^2+\frac{g_s}{2m}\vec\sigma\cdot\vec
    B\right)\psi_{\overline q}\,,
\]
where it should be noted that the covariant derivatives involve the generators $-T^{aT}$,
corresponding to those of anti-triplets.

The lowest order quark and antiquark Lagrangian in NRQCD is symmetric under
unitary transformations of the components of the Dirac spinors $\Psi_+$ and $X_-$ that preserve the
conditions $\frac12(1+\slashed u)\Psi_+=\Psi_+$ and $\frac12(1-\slashed u)X_-=X_-$. These
transformations form an internal symmetry group isomorphic to $SU(2)\times SU(2)$, and hence the heavy quark
spin component of total angular momentum is separately conserved, precisely as expected from
non-relativistic quantum mechanics. This is the additional spin-symmetry alluded to above.

The spin symmetry has considerable implications relating matrix elements of composite operators. An
important example is for the matrix elements of the four-fermion operators entering our
computation. For gamma-matrices $\Gamma$ and $\Gamma'$ spin symmetry implies
\begin{equation}
\label{eq:SpinSym1}
  \langle B_c|\overline\Psi_+^{(c)}\Gamma X_-^{(b)} \overline X_-^{(b)}\Gamma'\Psi_+^{(c)} |B_c\rangle
  =-\langle\langle Q\rangle\rangle
  \text{Tr}\left[ \Gamma \gamma_5 \left(\frac{1+\slashed{u}}2\right) \right]
   \text{Tr}\left[ \Gamma'     \left(\frac{1+\slashed{u}}2\right) \gamma_5 \right]\,,
 \end{equation}
where $\Psi_+^{(c)}$ and $X_-^{(b)}$ stand for the $\Psi_+$ and $X_-$ fields of the $c$ and $b$
quarks. This relates all of the matrix elements of  these 4-fermion operators to a single
invariant ``reduced matrix element'' $\langle\langle
Q\rangle\rangle$. It also gives
\begin{equation}
\label{eq:SpinSym2}
  \langle B_c|\overline X_-^{(b)}\Gamma_b X_-^{(b)} \overline\Psi_+^{(c)}\Gamma_c \Psi_+^{(c)}  |B_c\rangle
  =\langle\langle Q\rangle\rangle
  \text{Tr}\left[\gamma_5 \left(\frac{1+\slashed{u}}2\right) \Gamma_c    \left(\frac{1+\slashed{u}}2\right) \gamma_5 \Gamma_b\right] \,,
  \end{equation}
for arbitrary Dirac matrices  $\Gamma_c$ and $\Gamma_b$.

Higher order terms in the Lagrangian are readily incorporated. The first corrections to
$ \mathcal{L}_0$ are of order $v^7$  (relative order $v^2$).  Operators with powers
of $iu\cdot D$ are difficult to simulate on the lattice,  and therefore it is conventional (but unnecessary) to eliminate
them from the higher order corrections by means of field transformations. For example, the terms
\[
 \mathcal{L}=\mathcal{L}_0+\overline\Psi_+\left(-\frac1{4m^2}\left(i\slashed{D}_\perp (iu\cdot
     D)i\slashed{D}_\perp\right)
   +\frac1{8m^3}\left(i\slashed{D}_\perp (iu\cdot
     D)^2i\slashed{D}_\perp\right)\right)\Psi_++\cdots
\]
are eliminated in favour of the ones without temporal derivatives by the transformation
\begin{equation}
  \label{eq:field_transformation}
  \Psi_+\to\left(1+\frac{1}{8m^2}(i\slashed{D}_\perp)^2+\frac{1}{16m^3}\left((i\slashed{D}_\perp)^2iu\cdot
  D -2 (i\slashed{D}_\perp)(iu\cdot
  D) (i\slashed{D}_\perp)\right) \right)\Psi_+\;,
\end{equation}
which is chosen such that temporal derivatives in $\mathcal{L}$ appear
only in commutators with spatial derivatives, thus:
\begin{multline}
  \label{eq:lagm^3PreSimp}
 \mathcal{L}=\mathcal{L}_0+\overline\Psi_+\left(-\frac1{8m^2}\left([i\slashed{D}_\perp ,iu\cdot    D]i\slashed{D}_\perp+i\slashed{D}_\perp [iu\cdot    D,i\slashed{D}_\perp]\right)
   \right.\\\left.
   +\frac1{8m^3}\left((i\slashed{D}_\perp )^4-[i\slashed{D}_\perp ,iu\cdot    D]^2\right)\right)\Psi_++\cdots
\end{multline}
In order to use the result of this expansion in inverse powers of $m$
together with an expansion in powers of the relative velocity $v$, it is
necessary to display explicitly the (restframe) electric and magnetic
fields.  Using
$[i\slashed{D}_\perp ,iu\cdot D]=-ig_s\gamma^\mu u^\nu
G_{\mu\nu}$ (which contains only the chromo-electric
field $\vec E$ in the quarkonium restframe) and
$\slashed{D}_\perp\slashed{D}_\perp=D_\perp^2+\frac{1}{2}g_s\sigma^{\mu\nu}G_{\perp\mu\nu}$,
with
$G_{\perp\mu\nu}=(\delta^\lambda_\mu-u^\lambda u_\mu)
(\delta^\sigma_\nu-u^\sigma u_\nu)G_{\lambda\sigma}$ (the chromomagnetic
field $\vec B$ in the quarkonium restframe), this gives,
\begin{multline}
  \label{eq:lagm^3}
 \mathcal{L}=\mathcal{L}_0+\overline\Psi_+\bigg(-\frac{ig_s}{8m^2}\Big(
   u^\nu[i D_\perp^\mu G_{\mu\nu}]
     -i\sigma^{\mu\nu}u^\lambda(iD_{\perp\mu}G_{\nu \lambda}+ G_{\nu \lambda} iD_{\perp\mu}) \Big)
   \\
     +\frac1{8m^3}\Big((i{D}_\perp )^4 -\frac12g_s(i{D}_\perp)^2\sigma^{\mu\nu}G_{\perp\mu\nu}
     \\
     -\frac12g_s\sigma^{\mu\nu}G_{\perp\mu\nu}(i{D}_\perp )^2
       +\frac12g_s^2 G_{\perp\mu\nu}G_{\perp}^{\mu\nu}
     +i g_s^2\sigma^{\mu\nu}
       G_{\perp\mu\lambda}G_{\perp\nu}^{\lambda}\\
       -g_s^2u^\mu u^\nu   G_{\mu\lambda}G^{\lambda}{}_\nu
       -ig_s^2\sigma^{\mu\nu}u^\lambda u^\rho   G_{\mu\lambda}G_{\nu\rho}
     \Big)\bigg)\Psi_++\cdots
\end{multline}
where $[i D_\perp^\mu G_{\mu\nu}]$ indicates the derivative acting only on
the field strength tensor. The last line in \eqref{eq:lagm^3} is from
the square of the commutator, the last term in
\eqref{eq:lagm^3PreSimp}. Therefore, at order $v^7$
\begin{multline}
  \label{eq:Lagu^7}
    \mathcal{L}_1=\overline\Psi_+\bigg(c_1\frac1{8m^3} D_\perp^4-
    c_2\frac{g_s}{8m^2}u^\mu [D_\perp^\nu G_{\mu\nu}]
    -c_3\frac{g_s}{8m^2}i\sigma^{\mu\nu}u^\lambda(D_{\perp\mu}G_{\nu
      \lambda}+ G_{\nu \lambda} D_{\perp\mu}) \\
    - c_4\frac{g_s}{4m}\sigma^{\mu\nu}G_{\perp\mu\nu}
  \bigg)\Psi_+\,.
\end{multline}
The
dimensionless coefficients $c_n$, $n=1 \ldots,4$ are functions of $\alpha_s$ and are determined by
standard EFT
matching procedures.\footnote{See Ref.~\cite{Lepage:1992tx}. The
  formalism there is in terms of 2-component spinor fields. Below, when we
  quote results from the literature that uses 2-component spinor
  fields, we denote them by their
lowercase counterparts, $\Psi_+\to \psi_q$ and $X_-\to\chi$; with $\psi_{\bar q}=i\sigma^2 (\chi^\dag)^T$.} The tree-level calculation
above gives
$c_n=1+\mathcal{O}(\alpha_s)$, for all four coefficients. Beyond tree
level other operators may appear.  Using dimensional analysis, and
imposing symmetries, the additional possible operators can be
listed. Reparametrization invariance implies that the coefficient of
the second term in the Lagrangian in~\eqref{eq:Lagu^5} remains
unrenormalized (relative to the first term). It also imposes relations
on the coefficients of higher dimensional operators. In particular $c_1=1$ and
$c_3=2c_4-1$~\cite{Manohar:1997qy}.

At the next order in the expansion in inverse powers of $m$ one has
\[
 \mathcal{L}=\cdots+\overline\Psi_+\left(-
 \frac1{16m^4}\left(i\slashed{D}_\perp (iu\cdot
     D)^3i\slashed{D}_\perp\right)\right)\Psi_++\cdots
\]
from the expansion of the Lagrangian in Eq.~\eqref{eq:Lag-no-exp}, and various terms containing a single power of $iu\cdot
D$ and four powers of $i\slashed{D}_\perp$  that arise from the product of the lower order terms in
Eq.~\eqref{eq:Lag-no-exp} and those in the change of  field variables in
\eqref{eq:field_transformation}. Terms containing a single power of $iu\cdot D$ can be combined into
commutators so as to eliminate single temporal derivatives, $iu\cdot D$, at order $1/m^4$, provided
one further changes field variables via
\begin{equation}
  \label{eq:field_trans4}
   \Psi_+\to\left(1+\frac{1}{16m^4}(i\slashed{D}_\perp)^4 \right)\Psi_+\;.
 \end{equation}
 On the other hand, commutators of $D_\perp$ with $u\cdot D$ alone cannot remove all temporal
 derivatives from the Lagrangian at order $1/m^4$.\footnote{Although higher temporal derivatives are
   acceptable in effective field theories, as mentioned above, they are better avoided in lattice simulations
   of the quantum field theory. That these order $(i u\cdot D)^3$ terms cannot be removed may have
   escaped attention because they come in at order $v^{11}$,
 or relative to the lowest order, they constitute a
 relativistic correction
 of order $v^6$.}

The conserved Noether current associated with quark number that follows from $\mathcal{L}_0$ is
\begin{equation}
  \label{eq:conservedJ}
  J^\mu=u^\mu\overline\Psi_+\Psi_++\frac1{2m}\overline\Psi_+ i\leftrightvec{D}_\perp^\mu\Psi_+\,.
\end{equation}
Therefore one has
$\langle B_c(p')|u\cdot J|B_c(p)\rangle=\langle B_c(p')|\overline\Psi_+\Psi_+|B_c(p)\rangle$, since
$u\cdot D_\perp=0$. Then evaluating at $p'=p=M_{B_c}u$ and using spin symmetry\footnote{In the QCD case, defining the quark form factor $f(q^2)$ by
$\langle B_c(p')|j^\mu|B_c(p)\rangle=f(q^2)(p+p')^\mu$, where $j^\mu=\bar c\gamma^\mu c$ and
$q=p-p'$, charge conservation gives $f(0)=1$.}
\begin{equation}
  \label{eq:conservedDim3}
  \frac 1{2M_{B_c}}\langle
B_c(p)|\overline\Psi_+\Gamma\Psi_+ |B_c(p)\rangle= \frac12\text{Tr}\left[\Gamma\left(\frac{1+\slashed{u}}{2}\right)
\right]\,,
\end{equation}
and an analogous expression for the antiquark. This expression holds
non-perturbatively in $\alpha_s$ and receives corrections of order $v^2$ from the symmetry breaking
terms in $\mathcal{L}_1$. This is the basis for the statement above that quark
number conservation gives, non-perturbatively, the values of the matrix elements of the leading operators in the OPE that determines
$\Gamma(B_c\to X_{c\bar c})$ and $\Gamma(B_c\to X_{b\bar b})$. Furthermore we have
\[
\langle
B_c(p)|\overline\Psi_+ i\leftrightvec{D}_\perp^\mu\Psi_+ |B_c(p)\rangle=0\,.
\]
This has a simple physical interpretation: the right hand side is a
vector, but can only depend on the 4-velocity $u$, which however has
no perp-component (no spatial component in the $B_c$ rest-frame).

\section{Operator product expansion}\label{sec:OPE}

To obtain the lifetime of the $B_c$ meson, the optical theorem is used to relate the decay width to the forward scattering of the $B_c$ meson:
\begin{equation}
  \label{eq:GammaTran}
  \Gamma_{B_c} = \frac{1}{2 M_{B_c}}\langle B_c|\mathcal{T}|B_c\rangle\,,
\end{equation}
where the transition operator is given by the absorptive part of the time-ordered product:
\begin{equation}
  \label{eq:TranOpDef}
  \mathcal{T} = {\rm Im} \,i\int d^4x T \,\mathcal{H}_{eff}(x)\mathcal{H}_{eff}(0)\,.
\end{equation}
Invoking quark-hadron duality one expects that for a large energy release an OPE can be performed to
express the transition operator as a series of local operators of increasingly higher dimensions with
coefficients suppressed by the large energy released, corresponding in the case at hand to the heavy
quark masses, $m_q$. The calculation of the transition operator  is organized by separating the
contributions for $b$ and $c$ decays,  and those of WA and PI terms,
\[
  \mathcal{T}=\mathcal{T}_b+\mathcal{T}_c+\mathcal{T}_{\rm WA}+\mathcal{T}_{\rm PI}\,.
\]
Each of these is an expansion of operators, with $\mathcal{T}_{b,c}$ starting at dimension 3 while
$\mathcal{T}_{\rm WA, PI}$ starting at dimension 6. In NRQCD, these correspond to expansions
starting at order $v^3$ and $v^6$, respectively. Retaining the contributions
$\mathcal{T}_{\rm WA, PI}$ is physically important. It would appear then that for consistency we
need a full calculation to order $v^6$ which, since $\alpha_s\sim v$, should include corrections of
order $\alpha_s^3$ to the coefficients of the  dimension-3 operators. However, for WA
and PI the 2-body final state is enhanced  relative to the 3-body phase space of
single quark decay by $16\pi^2$ which is numerically $\sim\alpha_s(m_b)^{-3}$, and suppressed by the
probability that the quarks in the $B_c$ meet at a point, controlled by  the wave-function at the
origin, $|\Psi(0)|^2\propto f_{B_c}^2$, where $f_{B_c}$ is the $B_c$-meson decay constant.
Therefore we consider the expansions of $\mathcal{T}_{b,c}$ and $\mathcal{T}_{\rm WA, PI}$
independently and carry out each to some fixed order.

It must be noted that, since we use NRQCD to organize the calculation
in powers of the relative velocity $u$, the OPEs are reorganized. For example, the 4-fermion
operator $\overline X_-^{(b)}\Psi_+^{(c)} \overline\Psi_+^{(c)}   X_-^{(b)}$ and the magnetic moment
operator $ \overline\Psi_+^{(c)} \sigma_{\mu\nu}g_sG^{\mu\nu}_\perp\Psi_+^{(c)}$ are of mass
dimension 6 and 5 respectively but of order $v^6$ and $v^7$, respectively. We follow Ref.~\cite{Beneke:1996xe} in
including some terms of order $v^7$ in the expansion, merely to explore their significance. We do,
in contrast, retain all operators of order $v^6$ in the expansion of $\mathcal{T}_{\rm WA, PI}$. In
particular, we retain the operator  $ \overline\Psi_+^{(c)} \gamma_\mu D_\nu G^{\mu\nu}_\perp\Psi_+^{(c)}$
which we count as of order
$v^6$ since it is equivalent to a combination of 4-fermion penguin operators
$Q_3$-$Q_6$ which we keep in our calculations. The expansion of $\mathcal{T}_{b,c}$ is done consistently to order
$v^4$. The first non-perturbative effect in the calculation of $\mathcal{T}_{b,c}$ comes in at order
$v^5$ and we retain this together with the next  non-perturbative effect, of order $v^7$, to
explore the relevance of non-perturbative corrections. These corrections are particularly important
for charm decays, namely roughly of order 20\%. By comparison the omitted perturbative corrections are
expected to be of order $\alpha_s^2(m_c)\sim(0.37)^2$.

As inferred from the discussion above, we choose to expand in terms of operators in NRQCD. This has
several advantages, if only conceptually. First, it organizes the expansion systematically. To see
this consider that in the QCD expansion, an operator with $(\partial/m_q)^n q$, where $q$ is a
heavy quark, is not suppressed even though it carries arbitrary powers of the large mass in the
denominator. In NRQCD $i\partial/m_q$ is replaced by $v+i\partial/m_q $ the rest-frame velocity plus a
derivative that corresponds to the residual momentum, a small quantity by the NRQCD power
counting. Second, as we will see, the QCD expansion contains non-local operators, but not so the
NRQCD expansion. This is often ignored by writing Wilson coefficients as functions of the momenta,
$p_{b,c}$, of the $b,c$ quarks in the corresponding operators, but one should keep in mind that these
just stand for derivatives acting on the fields in the operators.  The Wilson coefficients often
contain negative powers of these momenta, that is, they are non-local operators. In NRQCD,
$p^{-2}=(m_qu+k)^{-2}$ has a local  expansion in terms of the derivative $k\to i\partial$. The third
advantage is that spin symmetry allows for vast simplifications which are absent in QCD. And finally,
if using NRQCD there is no non-trivial perturbative matching correction of the leading OPE operator,
as we now explain.

The perturbative matching calculation involves on-shell, or near on-shell
external quarks. The leading terms in the OPE consist of dimension 3 operators of the form
$\overline \Psi_+\Gamma\Psi_+$ (or the analogous anti-quark operators) with $\Gamma$ a Dirac
matrix. The matching calculation consists of evaluating the matrix element in  single quark states
$|\vec p.s\rangle$  of the transition operator in \eqref{eq:TranOpDef}
on the one hand, and the same matrix element of the OPE on the other, and
then fixing the Wilson coefficients by imposing equality  between the two calculations:
\[
  \langle \vec p, s|\mathcal{T}|\vec p, s\rangle = \sum_i C^{(3)}_i \langle \vec p, s|\overline
  \Psi_+\Gamma_i\Psi_+ |\vec p, s\rangle +\cdots
\]
In order to compute $C_i^{(3)}$ at $n$-th order in perturbation theory one must compute the matrix
elements on the right- and left-hand sides of this equation. However in NRQCD (but not in QCD) the
dimension 3 operators on the right-hand side are all protected from radiative corrections at zero
momentum transfer because they are related by spin symmetry to the conserved current
\eqref{eq:conservedJ} and just as in the case of mesons, the quark form factor has
$F(0)=1$.\footnote{That the electron form factor in QED satisfies $F(0)=1$ is clear from symmetry
  but non-trivial in perturbation theory.  Not only does it have to be carefully defined due to
  infrared divergences, but the perturbative series exponentiates. See \cite{Dahmen:1981ss} and
  references therein.} Incidentally, spin-symmetry allows us to treat the dimension 3 operators as a
single one, and this is standard practice: by choosing an operator in Eq.~\eqref{eq:conservedDim3}
that has a unit matrix element the imaginary part of the Wilson coefficient takes the value of the
perturbative quark decay width.

With this understanding we can now write
  \begin{equation}
  \label{eq:Tran}
  \mathcal{T}_{q} = C^{(3)}_q\bar Q Q + C^{(5)}_q\frac{1}{m_q^2}\bar Q \sigma_{\mu\nu}g_sG^{\mu\nu}Q+\sum_i C^{(6)}_{q,i} \frac{1}{m_q^3}O^{(6)}_i+\mathcal{O}(\frac1{m_q^4})\,.
\end{equation}
The computation
of the $C^{(i)}_q$ as well as that of $\mathcal{T}_{\rm WA, PI}$  will be discussed in the following subsections.

\subsection{$C_{c}^{(3)}$: Free $c$-quark decay}
\label{sec:Cc}
We turn first to the leading order term in the expansion in \eqref{eq:Tran}.
$C^{(3)}_{c}$ in Eq.~\eqref{eq:Tran} parametrizes the leading
contribution to the transition operator resulting from
$c$-quark decay.

Writing $Q(x)$ in terms of $\Psi_+(x)$ and performing the change of
field variables in \eqref{eq:field_transformation} and~\eqref{eq:field_trans4}, and retaining only
terms up to order $u^7$, one obtains
\begin{multline}
    \label{eq:QQexp}
\overline Q Q = \overline\Psi_+\Big(1+\tfrac1{2m^2} (iD_\perp)^2
  -\tfrac{g_s}{4m^2}\sigma^{\mu\nu}G_{\perp\mu\nu}
  +\tfrac{g_s}{4m^3}u^\nu[D_\perp^\mu G_{\mu\nu}]
       -\tfrac{ig_s}{4m^3}u^\nu\sigma^{\lambda\mu}(D_{\perp\lambda}G_{\mu\nu}+G_{\mu\nu}D_{\perp\lambda})
                                                                        \\
  -\tfrac{3}{16m^3}\big((iD_\perp)^2iu\cdot D+iu\cdot D(iD_\perp)^2\big)
         +\tfrac{13}{64m^4}(iD_\perp)^4+\cdots\Big)\Psi_+\,.
       \end{multline}
       This can be used for $\mathcal{T}_c$ and an analogous expansion in terms of $X_-$ for
$\mathcal{T}_b$.
When we need to evaluate matrix elements of the right hand side of Eq.~\eqref{eq:QQexp} we use
  the equations of motion to eliminate $iu\cdot D$ in favour of $-(iD_\perp)^2/2m$, we go to
  the restframe of the meson to decompose the field strength into chromo-electric and magnetic
  components and use 2-component spinors:
  \begin{align}
    \label{eq:nonRelExp-QQ-2com}
    \bar Q Q&\to \psi^\dag_q \Big(1+\tfrac1{2m^2} (\vec D)^2
    +\tfrac{g_s}{2m^2}\vec\sigma\cdot\vec B
    +\tfrac{g_s}{4m^3}[\vec D\cdot\vec E ]
         +\tfrac{ig_s}{4m^3}\vec\sigma\cdot(\vec D\times\vec E-\vec E\times\vec D)
                                                                 \nonumber         \\
        &\hspace{2.5in} +\tfrac{25}{64m^4}(\vec D)^4+\cdots\Big)\psi_q\,.
    \end{align}
 Below we neglect the spin-orbit coupling
  $\psi_q^\dagger g_s\vec{\sigma}\cdot(\vec{E}\times
  \vec{D}-\vec D\times\vec E)\psi_q$, since it has a vanishing
  matrix element.

The coefficient $C_c^{(3)}$  is given by the  decay rate of the charm
quark as if it were unbound,
\begin{equation}
  \Gamma_{c} = \Gamma_{c\to su\bar d}+\Gamma_{c\to su\bar s}+\Gamma_{c\to du\bar d}+\Gamma_{c\to du\bar s}+\Gamma_{c\to s\ell\nu}+\Gamma_{c\to d\ell\nu}\,.
\end{equation}
Throughout we approximate the light
quarks, $u$, $d$ as massless; as explained in
Sec.~\ref{sec:lightquarks} we compute both for $m_s=0$ and with a
non-vanishing $s$-quark pole mass given in terms of the running
$\overline{\text{MS}}$ mass.
The partial decay rate for $c\to su\bar d$ at $m_s=0$ and in the absence of running of Wilson
coefficients was first calculated by Guberina, Peccei and
R\"uckl \cite{Guberina:1979fe}. Altarelli, Curci, Martinelli and Petrarca first computed 2-loop
running and computed the 1-loop corrections to the decay rate anew using a subtraction
scheme common to both  calculations~\cite{Altarelli:1980fi}, and a
missing $\ln(\mu/m)$ term was corrected by Altarelli and Petrarca~\cite{Altarelli:1991dx}. Their result was confirmed by
Buchalla~\cite{Buchalla:1992gc}. Hokim and Pham computed the $\alpha_s$ corrections to the decay
rate for arbitrary final quark masses, including only the effect of $Q_2$ \cite{HOKIM1984202};
their results are trivially adapted to compute the $\alpha_s$ corrections to the decay rate if the only
operator were $Q_1$. The contribution to the rate from $Q_1$-$Q_2$ interference, for one massive
and two massless quarks in the final state, as is the case in, {\it e.g.,} $c\to s u\bar d$ and in $b\to c
\bar u d$, was given by Bagan, Ball, Braun and Gosdzinsky~\cite{Bagan:1994zd}.

Cabibbo  and Maiani computed the $\alpha_s$ corrections to the semileptonic decay neglecting the
charged lepton mass; the dependence on the final state quark mass was computed
numerically~\cite{Cabibbo:1978sw}. Finally, Nir gives $\Gamma_{\bar b\to \bar c
  \ell\nu}$, or equivalently, $\Gamma_{\bar c\to \bar s
  \ell\nu}$, for massless leptons~\cite{Nir:1989rm}. The
analytic expression for the 1-loop correction to the semileptonic rate can
be inferred  from the work of Hokim and Pham cited above. We note in
passing that, although not stated explicitly, it may be inferred that these works provide the rates
in terms of quark pole (on-shell) masses~\cite{Nir:1989rm}. We assume this is also the case for the
non-leptonic decay rates. As explained above, given a mass scheme one
has a perturbative expansion of the pole mass in terms of a well defined mass. Consistency requires
that when expressed in terms of the well defined mass, the rate can be expanded (and truncated)
to the appropriate order, {\it i.e.,} $\alpha_s^n$  with $n=2$ in the
Upsilon and meson schemes, and $n=1$ in the
$\overline{\text{MS}}$ scheme.
{We compute the quark decay rates at the one-loop order for the
  semi-leptonic decays and also for the contributions of the operators
  $Q_{1,2}$ in $\mathcal{H}_{eff}$ in the case of hadronic decays. For
  the penguin operators, $Q_{3-6}$,  however we only take compute the
  rate to leading order.}

Using the decay rate $\Gamma_c$ to compute the coefficient $C_c^{(3)}$
of $\bar Q Q$ in the OPE  is only appropriate for the
$\bar\Psi_+\Psi_+$ term in the expansion of $\bar Q Q$ in terms of
NRQCD fields, that is, the first term on the
right side of Eq.~\eqref{eq:QQexp}. The coefficients of each of the
remaining terms should be computed from individually matching them: in the
OPE one should really write
\begin{multline*}
  \mathcal{T}_q = C_q^{(3)}\overline\Psi_+\Psi_+ +  C_q^{(3,1)}\tfrac1{2m^2} \overline\Psi_+ (iD_\perp)^2 \Psi_+
  -\tfrac{g_s}{4m^2}C_q^{(3,2)}\overline\Psi_+\sigma^{\mu\nu}G_{\perp\mu\nu}\Psi_+
  \\
  +\tfrac{g_s}{4m^3}C_q^{(3,3)}\overline\Psi_+ u^\nu[D_\perp^\mu G_{\mu\nu}]\Psi_+
    -\tfrac{ig_s}{4m^3}C_q^{(3,4)}
  u^\nu\sigma^{\lambda\mu}(D_{\perp\lambda}G_{\mu\nu}+G_{\mu\nu}D_{\perp\lambda})
  +\cdots
\end{multline*}
where the ellipsis stand both for other terms in the matching of $\bar
Q Q$ and for the higher order terms in Eq.~\eqref{eq:Tran}.  At lowest
order $C_q^{(3,n)} = C_q^{(3)}$, with $n=1,2,\ldots$ for operators that
are present in the expansion \eqref{eq:QQexp} and vanish for operators
that arise in the $\bar Q Q$ expansion only at or above 1-loop
order. Reparametrization invariance requires that some of these operators come in fixed
  combinations, and this is reflected in exact relations between some of these coefficients, {\it
    e.g.}, $C_q^{(3,4)}=C_q^{(3,2)} $ \cite{Luke:1992cs}.

In our calculations we have used the leading order expression
for $\Gamma_q$ ($q=c,\bar b$) for the coefficients $C_q^{(3,n)}$, and
the next to leading expression for $\Gamma_q$ for $C_q^{(3)}$. Since
the subleading operators in \eqref{eq:QQexp} are of order $v^2$ and
higher, this truncation introduces uncertainties of order
$v^2\alpha_s\sim v^3$.

\subsection{$C_{\bar b}^{(3)}$: Free $\bar b$-quark decay}
The leading contribution to the transition operator resulting from the $\bar b$-quark stems from the coefficient $C^{(3)}_{\bar b}$ in Eq.~\eqref{eq:Tran}, which is given by the free anti-quark decay rate $\Gamma_{\bar b,spec}$ for the transition $\bar b\to \bar c$. The spectator decay rate takes the following form:
\begin{equation}
  \Gamma_{\bar b,spec} = \Gamma_{\bar b\to \bar c u\bar
    d}+\Gamma_{\bar b\to \bar c u\bar s}+\Gamma_{\bar b\to\bar c
    c \bar s}+\Gamma_{\bar b\to\bar c c\bar d}+\Gamma_{\bar b\to \bar c \ell\nu}+\Gamma_{\bar b\to \bar c \tau\nu}\,.
\end{equation}
Throughout we approximate the light
quarks, $u$, $d$ and $s$ as massless; the effect of a non-vanishing strange quark mass will also be
estimated.
The perturbative calculation to order $\alpha_s$ of individual (anti)-$b$ quark decay rates can be
found in the literature as detailed above, including the full $m_c$-dependence. In addition, Ref.~\cite{Bagan:1994qw}
gives $\Gamma_{\bar b\to \bar c \tau\nu}$ including both $m_c$ and $m_\tau$-dependence.
For the transition $\bar b\to \bar c u\bar d$ the expressions for the decay rate are given in
\cite{Bagan:1994zd}, who also estimated the case of a single massless and a massive pair of quarks
in the final state, which is the case for $b\to c\bar c s$. KLR corrected a misprint in the
latter\footnote{The full decay rate has first been reported in \cite{Bagan:1995yf}. As pointed out
  by KLR, there are several misprints in \cite{Bagan:1995yf}, resulting in a much lower decay rate
  for $b\to \bar c cs$. Also, KLR reports complex values for the contributions of vertex corrections
  to the decay rate. We have determined that the contribution to the rate is from the real part of
  those results ({\it i.e.}, not twice the real part, nor the imaginary part). We have verified this
  by an explicit recalculation of some of the vertex graphs and the corresponding real emission
  graphs. In addition we have verified that with this interpretation the resulting width has a
  finite limit as the gluon (IR regulator) mass approaches zero. }  and included additional effects,
most importantly that of penguin operators and of writing the rate in terms of a well defined
mass. As in KLR, we  treat the penguin Wilson coefficients as of sub-leading
  order; while this is not formally correct, it is practical, since the full calculation of the
  decay rate at 1-loop is unavailable. We also follow KLR in including also
$Q_{1,2}$ contributions to the rate through ``penguin diagrams'', and the contribution of the
chromomagnetic operator, $\bar b\sigma^{\mu\nu}G^A_{\mu\nu}T^A b$.

\subsection{$C_{\bar b}^{(5)}$: Chromomagnetic operator $O_{b8}$}
The coefficient $C_{\bar b}^{(5)}$ governs the contribution of the
chromomagnetic operator
\begin{equation}
  O_{b8} = (\overline{X}_-^{(b)} \sigma^{\mu\nu} g_s T^A  X_-^{(b)}) G^{A}_{\mu\nu}\,,
\end{equation}
to the total decay rate $\Gamma_{B_c}$.

Writing $Q(x)$ in terms of $\Psi_+(x)$ and performing the change of
field variables in \eqref{eq:field_transformation} and~\eqref{eq:field_trans4}, and retaining only
terms up to order $v^7$, one obtains
 \begin{equation}
     \label{eq:QsigmaQexp}
           \overline Q \sigma^{\mu\nu}G_{\mu\nu}Q = \overline\Psi_+\Big(\sigma^{\mu\nu}G_{\perp\mu\nu}
           -\frac{1}{m}\left(u_\nu[D_{\perp\mu}G^{\mu\nu}]
                +i\epsilon^{\mu\nu\lambda\rho}G_{\mu\nu}D_{\perp\lambda}\gamma_\rho\gamma_5\right)
           +\cdots  \Big)\Psi_+\,.
  \end{equation}
This can be used for $\mathcal{T}_c$ and an analogous expansion in terms of $X_-$ for
$\mathcal{T}_b$.
When we need to evaluate matrix elements of the right-hand side of Eq.~\eqref{eq:QsigmaQexp} we go to
  the restframe of the meson to decompose the field strength into chromo-electric and magnetic
  components and use 2-component spinors:
  \begin{equation}
    \label{eq:nonRelExp-QsigmaQ-2com}
        \bar Q \sigma_{\mu\nu}g_sG^{\mu\nu} Q \to \psi_q^\dagger \left(    -2 g_s\vec\sigma\cdot\vec B
                                             -\frac{g_s}{m}[\vec D\cdot\vec
                                             E]-\frac{ig_s}{m}\vec\sigma\cdot(\vec D\times\vec E-\vec E\times\vec D)\right)\psi_q\,.
    \end{equation}
The
  chromomagnetic moment operator, with coefficient $C_q^{(5)}$ above
  is of order $v^7$ in the NR expansion, so to this order there are no
  additional operators resulting from the field redefinition. An
  electromagnetic moment operator has been ignored since it is further
  suppressed by the smallness of the fine structure constant.

The coefficient
$C_{\bar b}^{(5)}$ consists of several contributions:
\begin{multline}
  C_{\bar b}^{(5)} = -\Gamma_{0b}\big[P_{ce\nu}+P_{c\mu\nu} + P_{c\tau\nu} +N_a(\mu_b) (P_{cu(s+d)1}+P_{cc(s+d)1})\\
+ N_b(\mu_b) (P_{cu(s+d)2}+P_{cc(s+d)2})\big]\,,
\end{multline}
where the normalization constant
\begin{equation}
  \Gamma_{0b} = \frac{G_F^2m_b^5}{192\pi^3}|V_{cb}|^2\,,
\end{equation}
sets the scale for tree-level decay rate of the
$\bar b$ quark. The subscript of the ``phase space'' factors\footnote{The so-called phase space
  factors are really integrals over phase space of the square modulus of amplitudes that do not have
  trivial dependence on kinematic variables.}  $P_i$ denote the particles in the loop and $N_i$ denote the following combinations of Wilson coefficients:
\begin{align}\label{eq:NaNb}
  N_a(\mu) &= 3C_1(\mu)^2+3C_2(\mu)^2+2C_1(\mu)C_2(\mu)\,, \\\notag
  N_b(\mu) &= 8C_1(\mu)C_2(\mu)\,.
\end{align}
Neglecting all fermion masses except for the charm and tau mass one finds the following phase space factors \cite{Beneke:1996xe}. The light contributions are given by
\begin{align}\label{eq:Pc1}
  P_{cu(s+d)1}=P_{ce\nu}=P_{c\mu\nu}&= (1-x_c)^4\,,\\
  P_{cu(s+d)2}&= (1-x_c)^3\,,
\end{align}
with $x_c = (m_c/m_b)^2$. The semi-leptonic mode is given by
\begin{align}\label{eq:Pctaunu}
  P_{c\tau\nu} &= \sqrt{1-2(x_\tau+x_c)+(x_\tau-x_c)^2}\left[1-3(x_\tau+x_c)+3(x_\tau^2+x_c^2)\right.\notag\\
  &\left.\qquad-x_\tau^3-x_c^3-4x_\tau x_c+7x_\tau x_c(x_\tau+x_c)\right]\\\notag
  &\quad+12 x_\tau^2x_c^2 \ln\left(\frac{\left(1-x_\tau-x_c+\sqrt{1-2(x_\tau+x_c)+(x_\tau-x_c)^2}\right)^2}{4x_\tau x_c}\right)\,,
\end{align}
with $x_\tau = (m_\tau/m_b)^2$. In the limit of massless tau leptons
$(x_\tau\to 0)$ Eq.~\eqref{eq:Pctaunu} reduces to Eq.~\eqref{eq:Pc1}.
The contributions from $Q_1$ and $Q_2$ are given by
\begin{equation}
  P_{cc(s+d)1} = \sqrt{1-4x_c}(1-6x_c+2x_c^2+12x_c^3)+24x_c^4\ln\left(\frac{1+\sqrt{1-4x_c}}{1-\sqrt{1-4x_c}}\right)\,,
\end{equation}
and for the second insertion by
\begin{equation}\label{eq:Pc2}
  P_{cc(s+d)2} = \sqrt{1-4x_c}(1+\frac{x_c}{2}+3x_c^2)-3x_c(1-2x_c^2)\ln\left(\frac{1+\sqrt{1-4x_c}}{1-\sqrt{1-4x_c}}\right)\,.
\end{equation}
Again, these expressions are given as a function of pole masses, which have to be written in terms of
well defined masses. At this level we only have expressions at zeroth order in $\alpha_s$, and for
consistency we truncate the expansion of pole masses at zeroth order as well, {\it e.g.},
$m_b=m_\Upsilon/2$ in the Upsilon scheme.

\subsection{$C_{c}^{(5)}$: Chromomagnetic operator $O_{c8}$}
The $c$-quark analogue of the chromomagnetic operator involving $b$-quarks  is given by
\begin{equation}
  O_{c8} = (\overline\Psi_+^{(c)} \sigma^{\mu\nu} g_s T^A\Psi_+^{(c)}) G^{A}_{\mu\nu}\,,
\end{equation}
and it contributes in the following way to the transition operator $\mathcal{T}_c$:\footnote{In this expression we neglect the Cabibbo suppressed mode but include it in the numerical analysis for $m_s\neq 0$.}
\begin{equation}
  C_{c}^{(5)} = -\Gamma_{0c}\left[P_{se\nu} + P_{s\mu\nu} +|V_{ud}|^2(N_a(\mu_c) P_{sud1} + N_b(\mu_c) P_{sud2})\right]\,,
\end{equation}
with the tree-level decay rate

\begin{equation}
  \Gamma_{0c}=\frac{G_F^2m_c^5}{192\pi^3}|V_{cs}|^2\,,
\end{equation}
and the phase space factors \cite{Beneke:1996xe}
\begin{align}
  P_{se\nu} = P_{s\mu\nu}= P_{sud1} &= (1-x_s)^4\,,\\
  P_{sud2} &= (1-x_s)^3\,,
\end{align}
with $x_s=(m_s/m_b)^2$. The Wilson coefficient combinations are defined in Eq.~\eqref{eq:NaNb} but we
have indicated that they are evaluated at a scale $\mu_c$ that may be different than for
$C_b^{(5)}$, since the choice of charm scale $\mu_c\sim \mathcal{O}(m_c)$ is more appropriate.

\subsection{Pauli interference}
For the dimension-six contributions to the transition operator in Eq.~\eqref{eq:Tran} we write, as
is customary, the full contribution coming from the operator insertions rather than the individual
coefficients $(C_q^{(6)})_i$. This allows for a more compact presentation of the results~\cite{Beneke:1996xe, Neubert:1996we, Chang:2000ac}:
\begin{multline}
  \label{eq:TansitionPI}
  \mathcal{T}_{\text{PI}}  =
  \frac{4G_F^2}{\pi}|V_{cb}|^2p_-^2(1-z_-)^2
  \bigg\{ (C_1+C_3)(C_2+ C_4)(\bar b^\alpha \gamma_\mu P_L b^\beta)(\bar c^\beta \gamma^\mu P_L c^\alpha) \\
    +\frac12\big((C_1+C_3)^2+(C_2+ C_4)^2\big)(\bar b \gamma_\mu P_L b)(\bar c \gamma^\mu P_L c)\\
   -\frac{m_c}{4p_-^2}\bigg[\Big((C_1+C_3)C_5+(C_2+C_4)C_6\Big)
   \Big((\bar b \gamma_\mu P_L b)(\bar c\slashed{p}_- \gamma^\mu P_L c)+(\bar b \gamma_\mu P_L b)(\bar c \gamma^\mu \slashed{p}_-P_R c)\Big) \\
   +\Big((C_1+C_3)C_6+(C_2+C_4)C_5\Big)
   \Big((\bar b^\alpha \gamma_\mu P_L b^\beta)(\bar c^\beta \slashed{p}_- \gamma^\mu P_L c^\alpha)+(\bar b^\alpha \gamma_\mu P_L b^\beta)(\bar c^\beta \gamma^\mu \slashed{p}_-P_R c^\alpha)\Big)\bigg] \\
   +\bigg[\frac{(1-z_-)}{12}g^{\mu\nu}+\left(\frac{1}{2}-\frac{(1-z_-)}{3}\right)
     \frac{p_-^\mu p_-^\nu}{p_-^2}\bigg]\times\\
 \Big((C_5^2+C_6^2)(\bar b \gamma_\mu P_L b)(\bar c \gamma_\nu P_R c)+2C_5C_6(\bar b^\alpha \gamma_\mu P_L b^\beta)(\bar c^\beta \gamma_\nu P_R c^\alpha)\Big)\bigg\}\,,
\end{multline}
where  $p_-=p_b-p_c$ and $z_-=m_c^2/p_-^2$. We have summed over $s$ and $d$ quarks, neglected their
masses and used  $|V_{cd}|^2+|V_{cs}|^2\approx 1$.

The notation  $p_-=p_b-p_c$ should be understood in the
operator sense, that is, as derivatives acting on the $b$ and $c$ fields. Recall that we  have
chosen to match the OPE
directly to NRQCD; by a slight abuse of notation  we have denoted the fields in
Eq.~\eqref{eq:TansitionPI} as in the full theory, but it should be understood that they really stand for
EFT fields, {\it i.e.}, $c\to \Psi_+^{(c)}$ and $b\to X_-^{(b)}$. One advantage of matching directly
to the EFT is that $p_-=(m_b-m_c)u+(k_b-k_c)$ where the derivatives $k_b-k_c$ correspond to residual
momenta, and are clearly sub-leading in the NR expansion. While some authors use
$p_-^2\approx2(m_b^2+m_c^2)-M_{B_c}^2$ \cite{Chang:2000ac}, in our  approach the leading term in the
NR expansion is unambiguous and the higher order terms correspond to matrix elements of well defined
operators.

Another advantage of matching to the effective theory is that one may then avoid non-local operators
in the expansion. Note that there are inverse powers of $p_-^2$ implicit in the functions of $z_-$ in
Eq.~\eqref{eq:TansitionPI}. By matching to the EFT the inverse powers give well defined expansions
in terms of local operators, $1/p_-^2=1/(m_b-m_c)^2[1-2u\cdot(k_b-k_c)/(m_b-m_c)+\cdots]$.

\subsection{Weak annihilation}
Also for the WA contributions we will report the full contribution to the transition operator instead of just the coefficients $(C_q^{(6)})_i$. The semi-leptonic contribution to the transition operator reads \cite{Beneke:1996xe}:
\begin{equation}
\label{eq:TansitionWASL}
  \mathcal{T}^{\text{SL}}_{\text{WA}} = -\frac{4G_F^2}{\pi}|V_{cb}|^2p_+^2\left[\frac{(1-z_\tau)^2}{12}g^{\mu\nu}+\left(\frac{(1-z_\tau)^2}{2}-\frac{(1-z_\tau)^3}{3}\right)\frac{p_+^\mu p_+^\nu}{p_+^2}\right](\bar b^\alpha \gamma_\mu P_L b^\beta)(\bar c^\beta \gamma_\nu P_L c^\alpha)\,,
\end{equation}
with $p_+=p_b+p_c$ and $z_\tau=m_\tau^2/p_+^2$. As was the case for $p_-$ in the PI
computation, the notation  $p_+=p_b+p_c$ here should be understood in the
operator sense, that is, as derivatives acting on the $b$ and $c$ fields. In addition, since we  have
chosen to match the OPE
directly to NRQCD the fields indicated should be those for NRQCD; by a slight abuse of notation, and
in order to make the long expression more legible, we have denoted the fields in
Eq.~\eqref{eq:TansitionWASL} and below in Eq.~\eqref{eq:TansitionWA}  as those of  the full theory, but it should be understood that they really stand for
 fields in the effective theory, {\it i.e.}, $c\to \Psi_+^{(c)}$ and $b\to X_-^{(b)}$. One advantage of matching directly
to the EFT is that $p_+=(m_b+m_c)u+(k_b+k_c)$ where the derivatives in $k_b+k_c$ correspond to residual
momenta, and are clearly sub-leading in the NR expansion, and the leading term is well defined in
terms of pole masses and the meson 4-velocity.

The hadronic WA decay gives
\begin{multline}
  \label{eq:TansitionWA}
  \mathcal{T}_{\text{WA}} =
                                        -\frac{4G_F^2|V_{cb}|^2}{\pi}
                                         p_+^2(1-z_+)^2\times\\
    \bigg\{\left[\frac{(1-z_+)}{12}g^{\mu\nu}+\left(\frac{1}{2}-\frac{(1-z_+)}{3}\right)\frac{p_+^\mu p_+^\nu}{p_+^2}\right]\Big[(C_1+C_3)^2(\bar b \gamma_\mu P_L b)(\bar c \gamma_\nu P_L c)  \\
  +\Big(2(C_1+C_3)(C_2+C_4)+N_c (C_2+C_4)^2\Big)(\bar b^\alpha \gamma_\mu P_L b^\beta)(\bar c^\beta \gamma_\nu P_L c^\alpha)\Big] \\
    -\frac{m_c}{4p_+^2}\bigg[(C_1+C_3)C_5\Big((\bar b\slashed{p}_+P_Lb)(\bar cc)+i(\bar b\gamma^\mu P_L b)(\bar c\sigma_{\mu\nu}p_+^\nu\gamma_5c)\Big)\\
+\Big(C_6(C_1+C_3)+C_5(C_2+C_4)+N_cC_6(C_2+C_4)\Big)\times \\
\hspace{1cm}\Big((\bar b^\alpha\slashed{p}_+P_Lb^\beta )(\bar c^\beta
    c^\alpha )+i(\bar b^\alpha \gamma^\mu P_L b^\beta)(\bar c^\beta
    \sigma_{\mu\nu}p_+^\nu\gamma_5c^\alpha)\Big)\bigg] \\
+\tfrac12\Big(C_5^2(\bar b \gamma_\mu P_L b)(\bar c   \gamma^\mu P_R c)
  +(N_cC_6^2+2C_5C_6)(\bar b^\alpha \gamma_\mu P_L b^\beta)(\bar c^\beta \gamma^\mu P_R c^\alpha) \Big)\bigg\}\,,
\end{multline}
with $z_+=m_c^2/p_+^2$. We have summed over $s$ and $d$ quarks, neglected their masses and used $|V_{cd}|^2+|V_{cs}|^2\approx 1$.

\section{Matrix elements}
\label{sec:MEs}

As mentioned above, we denote 2-component spinor fields by the
lowercase counterparts of their 4-component Dirac forebearers, $\Psi_+\to \psi_q$ and $X_-\to\chi$ and $\psi_{\bar q}=i\sigma^2 (\chi^\dag)^T$.
The first correction in Eq.~\eqref{eq:Tran} can be estimated using potential
models~\cite{Kiselev:1994rc}
\begin{equation}\label{eq:ME-T}
  \frac{\langle B_c| \psi^\dag_c (i\vec{D})^2\psi_c^{\phantom{\dag}}|B_c\rangle}{2M_{B_c} } = \frac{\langle B_c|
    \psi^\dag_{\overline b} (i\vec{D})^2\psi_{\overline b}^{\phantom{\dag}}|B_c\rangle}{2M_{B_c} }  = \frac{2m_cm_b}{(m_c+m_b)}T \,,
\end{equation}
where $T$ is the expectation value of the kinetic energy computed in potential models. Note that the
first equality above is interpreted as $(m_c v_c)^2 =(m_b v_b)^2$ which is useful in estimating the
NR-quark velocities, $v_b$ and $v_c$.
For our calculations we estimate the matrix element of $\vec D^4$ as the square of that of
  $\vec D^2$, from  Eq.~(\ref{eq:ME-T}), thus:
\begin{equation}
  \frac{\langle B_c| \psi^\dag_c (i\vec{D})^4\psi_c^{\phantom{\dag}}|B_c\rangle}{2M_{B_c} } = \frac{\langle B_c|
    \psi^\dag_{\overline b} (i\vec{D})^4\psi_{\overline b}^{\phantom{\dag}}|B_c\rangle}{2M_{B_c} }  = \frac{4m_c^2m_b^2}{(m_c+m_b)^2}T^2 \,.
\end{equation}

The leading matrix elements for the chromomagnetic operator are given by
\begin{align}
  \frac{\langle B_c| \psi^\dag_{\overline b} g_s\vec{\sigma}\cdot \vec{B}\psi_{\overline b}^{\phantom{\dag}}|B_c\rangle}{2M_{B_c} } &= -\frac{4}{3}g_s^2\frac{|\Psi(0)|^2}{m_c} \,, \\
    \frac{\langle B_c| \psi^\dag_{\overline b} g_s[\vec{D} \cdot \vec{E}]\psi_{\overline b}^{\phantom{\dag}}|B_c\rangle}{2M_{B_c} } &= \frac{4}{3}g_s^2|\Psi(0)|^2 \,,
\end{align}
and the corresponding matrix elements for the charm quark are obtained by the replacement
$m_b\leftrightarrow m_c$. The wave function at the origin, $\Psi(0)$, relates these to other
physical quantities that can be determined from Monte Carlo simulations of NRQCD on the lattice:
\begin{equation}
  f_{B_c}^2=\frac{12|\Psi(0)|^2}{M_{B_c}}\,,\qquad M_{B_c^*}-M_{B_c} =\frac{8}{9}g_s^2\frac{|\Psi(0)|^2}{m_b m_c}\,.
\end{equation}

The matrix elements of four quark operators are all related by spin symmetry per
Eqs.~\eqref{eq:SpinSym1} and~\eqref{eq:SpinSym2}. We have
\begin{align}
    \langle B_c | (\overline X_-^{(b)\alpha} \gamma_\mu P_L  X_-^{(b)\beta})(\overline\Psi_+^{(c)\beta} \gamma_\nu P_L \Psi_+^{(c)\alpha}) | B_c \rangle &= \frac{f_{B_c}^2B_{B_c}}{4}\left(\frac{1}{2}q^2g^{\mu\nu}-q^\mu q^\nu\right) \,,\\\notag
        \langle B_c | (\overline X_-^{(b)}  \gamma_\mu P_L X_-^{(b)})(\overline\Psi_+^{(c)} \gamma_\nu P_L \Psi_+^{(c)}) | B_c \rangle   &= \frac{f_{B_c}^2 B'_{B_c}}{12}\left(\frac{1}{2}q^2g^{\mu\nu}-q^\mu q^\nu\right) \,,
\end{align}
where $q$ is the  momentum of the $B_c$. If $  \langle B_c |
  (\overline X_-^{(b)}  T^A \gamma_\mu P_L
  X_-^{(b)})(\overline\Psi_+^{(c)} T^A\gamma_\nu P_L \Psi_+^{(c)}) |
  B_c \rangle = 0 $ then $B'_{B_c}=B_{B_c}$. Using spin symmetry, we find for
  the new matrix elements for the penguin operators  that enter the
  calculation of PI:
\begin{align}
  \langle B_c | (\overline X_-^{(b)} \gamma_\mu P_L  X_-^{(b)} )(\overline\Psi_+^{(c)}\gamma^\nu  \gamma^\mu P_L \Psi_+^{(c)}) | B_c \rangle
                  &=\frac{f_{B_c}^2B'_{B_c}}{12}M_{B_c} q^\nu \,,\\\notag
  \langle B_c | (\overline X_-^{(b)}  \gamma_\mu P_L  X_-^{(b)} )(\overline\Psi_+^{(c)} \gamma^\mu \gamma^\nu P_R \Psi_+^{(c)}) | B_c \rangle
                  &= \frac{f_{B_c}^2B'_{B_c}}{12}M_{B_c}q^\nu \,,\\\notag
\langle B_c | (\overline X_-^{(b)}  \gamma_\mu P_L  X_-^{(b)} )(\overline\Psi_+^{(c)} \gamma_\nu P_R \Psi_+^{(c)}) | B_c  \rangle
                  &= -\frac{f_{B_c}^2B'_{B_c}}{24}M_{B_c}^2g_{\mu\nu} \,.
\end{align}
The colour-crossed matrix elements are obtained from these by
  replacing $B'_{B_c}\to 3B_{B_c}$.
For WA the additional matrix elements are given by:
\begin{align}
  \langle B_c | (\overline X_-^{(b)} \gamma^\mu P_L X_-^{(b)} )(\overline\Psi_+^{(c)}\Psi_+^{(c)}) | B_c \rangle &=-\frac{f_{B_c}^2B'_{B_c}}{12}M_{B_c} q^\nu \,,\\\notag
\langle B_c | (\overline X_-^{(b)} \gamma^\mu P_L  X_-^{(b)}
  )(\overline\Psi_+^{(c)}\sigma_{\mu\nu}\gamma_5\Psi_+^{(c)}) | B_c \rangle &=
                                                 i\frac{f_{B_c}^2B'_{B_c}}{4}M_{B_c} q_\nu \,,
\end{align}
The ``bag parameters'' $B_{B_c}$ and $B_{B_c}'$ have been chosen so that $B_{B_c}=B_{B_c}'=1$ in the vacuum insertion
approximation.\footnote{We hasten to indicate that the many bag parameters used in
  Ref.~\cite{Chang:2000ac} to characterize various matrix elements are not independent at this
  order in the NR-expansion because of the spin symmetry relations, given in Eqs.~\eqref{eq:SpinSym1}
  and~\eqref{eq:SpinSym2}.} The vacuum insertion approximation can be justified in the large $N_c$ limit, so the errors incurred in using these
expressions with $B_{B_c}=B_{B_c}'=1$ are of the order
$\mathcal{O}(1/N_c)$ and
$\mathcal{O}(v)$.\footnote{Ref.~\cite{Beneke:1996xe} states that large $N_c$ is not required because ``deviations from factorization arise from higher Fock components of the
  $B_c$  wavefunction''.} For the numerical estimates below we adopt this approximation and then
analyze the error incurred in the computation of the lifetime by varying $B_{B_c}$ and $B'_{B_c}$
away form unity. Eventually, calculations of these matrix elements in Monte Carlo
  simulations of NRQCD on the lattice will remove this uncertainty.
In terms of these parameters WA and PI contributions to the width are
obtained from the matrix elements of the transition operators in Eqs.~\eqref{eq:TansitionWA}
and~\eqref{eq:TansitionPI}, respectively, yielding:
\begin{multline}
  \label{eq:GammaWA}
  \Gamma^{\text{WA}}=
 \frac{G_F^2 f_{B_c}^2  M_{B_c}  } {2
   \pi } |V_{cb}|^2 (1-z_+)^2  ( {m_b}+ {m_c})^2\times\\
   \left\{z_+\left[(C_1+C_3)^2\frac{B_{B_c}'}{12}+\Big(2(C_1+C_3)(C_2+C_4)+3(C_2+C_4)^2\Big)\frac{B_{B_c}}{4}\right]\right.\\
  \left. -\frac{m_c}{m_b+m_c}\left[ (C_1+C_3)C_5\frac{B_{B_c}'}{3}+\Big(C_6(C_1+C_3)+(C_2+C_4)(C_5+3C_6)\Big)B_{B_c}\right]\right.\\
  \left.+C_5^2\frac{B_{B_c}'}{3}+(3C_6^2+2C_5C_6)B_{B_c}\right\}\,,
\end{multline}
and
\begin{multline}
    \Gamma^{\text{PI}}=\frac{f_{B_c}^2  G_F^2 M_{B_c}}{4 \pi } |V_{cb}|^2 (1-z_-)^2 ({m_b}-{m_c})^2\times\\
    \bigg\{2 ( {C_1}+ {C_3}) ( {C_2}+ {C_4})B_{B_c}+\Big( ( {C_1}+ {C_3})^2+ ( {C_2}+ {C_4})^2\Big) \frac{B_{B_c}'}{3} \\
   -\frac{m_c}{m_b-m_c}\left[\Big(( {C_1}+ {C_3})C_5+ ({C_2}+ {C_4})C_6 \Big) \frac{B_{B_c}'}{3} +\Big (( {C_1}+ {C_3})C_6+ ({C_2}+ {C_4})C_5 \Big)B_{B_c}\right]\\
  -(C_5^2+C_6^2) \frac{B_{B_c}'}{6} -C_5C_6B_{B_c}
   \bigg\}\,,
  \end{multline}
where $z_\pm=(m_b/m_c\pm1)^{-2}$.

\section{Numerical Analysis}\label{sec:num}

In this section we present the results for the $B_c$ decay width in
the $\overline{\text{MS}}$, the meson and the Upsilon scheme.  For our
analysis we use the input values summarized in
Tab.~\ref{tab:num-input}. The matrix elements of the relevant operators are determined
using potential models and spin-symmetry (see
Sec.~\ref{sec:MEs}). The $\overline{\text{MS}}$ masses are the averages of the two most recent
and precise lattice calculations:
For $\overline{m}_b$ we average the results of \cite{Bazavov:2018omf} and \cite{Colquhoun:2014ica} and find
\[\overline{m}_b(\overline{m}_b) = 4.195(9)\,\text{GeV}\,,\]
and for $\overline{m}_c$ using \cite{Bazavov:2018omf,Lytle:2018evc}
we obtain
\[\overline{m}_c(\overline{m}_c) = 1.2734(44)\,\text{GeV}\,.\]

The QCD coupling constant in the QCD
corrections are calculated using the 1-loop beta function. As
explained in previous sections, the QCD corrections are carried out to
1st order in $\alpha_s$, with the Wilson coefficients computed to NLL
order. The running of the Wilson coefficients $C_{1,2}$ incorporates
analytically the effect of the 2-loop beta function, while that of
$C_{3-6}$ is computed only at LL; since $C_{3-6}$ are very small, the
numerical effect of this approximation is negligible in the total
rate.  As is well known, the consistent counting for resummation of
logs involves 2-loop beta functions and anomalous dimensions in the
running and 1-loop matching and matrix elements. Hence it is
appropriate to include only a 1-loop
running $\alpha_s(\mu)$ in the matrix elements. In particular,
including the effects of 2-loop running in the matrix elements only
increases the $\mu$ dependence of the final results for partial and
total decay widths. The renormalization scale $\mu$ has been chosen differently for
different partial widths. In calculations of $\bar b$ decays and for
WA and PI we use $\mu=\overline m_b(\overline m_b)$, while for $c$-decays we use
$\mu=\overline m_c(\overline m_c)$.\footnote{In BB the WA and PI contributions were evaluated at an intermediate scale $\mu_{3}=2 \frac{m_b m_c}{m_b + m_c}$.}

For the semileptonic $\bar b$-decays and for WA the
$m_\tau$-dependence is taken into account, whereas the light leptons
are assumed to be massless. Furthermore we neglect the light quark masses. In
particular, the  strange mass, $m_s$, is neglected in
the $\bar b$-decays.

It is worth repeating that for the computation we take into account
QCD corrections truncated to order $\alpha_s$ and carry out the non-relativistic expansion up to $v^7$ (relative order $v^4$ since the leading order is
$v^3$) as presented in the previous sections. Since the
power counting in NRQCD has $v\sim\alpha_s$ this is not fully
consistent. However, the numerically important effects of WA and PI come in first at order
$v^6$. Roughly, these effects are amplified by a factor of $16\pi^2$
from the 2-body vs 3-body decay phase space, and suppressed by a
relative order $v^3$
factor of $(f_{B_c}/M_{B_c})^2$, and  $16\pi^2
(f_{B_c}/M_{B_c})^2=0.73$. Corrections to 3-body decays of relative
order $v^2$ and $v^4$ are included, so their numerical effect can be
estimated and analyzed. We do not
include any QCD  corrections to WA and PI, since these would correspond
to higher-order velocity terms which are neglected in our counting.

\begin{table}[t]
\centering
\renewcommand{\arraystretch}{1.3}
\resizebox{\columnwidth}{!}{
\begin{tabular}{|llllll|}
\hline
  Parameter
& Value
& Ref.
&  Parameter
& Value
& Ref.
\\
\hline\hline
  $G_F$                   & $1.166379 \times 10^{-5} \GeV^{-2}$  & \cite{Tanabashi:2018oca}
& $\alpha_s(M_Z)$ & $0.1179\pm 0.0010$ & \cite{Tanabashi:2018oca}
\\
\hline
$|V_{cb}|$               & $0.0410(14)$         & \cite{Tanabashi:2018oca}
& $|V_{ud}|$               & $0.97370$          &
\\
$|V_{cs}|$               & $0.97320(11)$          & \cite{Tanabashi:2018oca}
&  $|V_{cd}|$                 & $0.22636$  &
\\
\hline
$M_W$     & $80.385$ GeV  &
& $M_Z$                   & $91.1876$ GeV         &
\\
  $M_{B_c}$     & $6274.9\pm 0.8$ MeV  & \cite{Tanabashi:2018oca}
& $f_{B_c}$                   & $0.427(6)$ GeV         & \cite{McNeile:2012qf}
\\
  $M_{B_c^*}-M_{B_c}$                 & $54(3)$ MeV     & \cite{Dowdall:2012ab}
& $\overline m_c(\overline m_c)$            &  $1.2734(44)$ GeV       & \cite{Bazavov:2018omf,Lytle:2018evc}
\\
  $M_{\Upsilon}(1S)$             & $9460.30(26)$ MeV           & \cite{Tanabashi:2018oca}
& $\overline m_b(\overline m_b)$            & $4.195(9)$ GeV      & \cite{Bazavov:2018omf,Colquhoun:2014ica}
\\
$M_{J/\Psi}(1S)$             & $3096.900(6)$ MeV           & \cite{Tanabashi:2018oca}
            &  $\overline m_s(2\GeV)$      &$93^{+11}_{-5}$ MeV & \cite{Tanabashi:2018oca}
\\
    $m_b/m_c$        &      4.577$\pm$0.008      &  \cite{Bazavov:2018omf}
&      $T$      &    0.37$\pm$0.04 GeV   & \cite{Kiselev:1994rc}
\\
$m_\tau$        &      $1776.86\pm 0.12$ MeV      &  \cite{Tanabashi:2018oca}
&     $\lambda_1$       &  $-0.27\pm 0.14$    & \cite{Hoang:1998ng}
\\
\hline
\end{tabular}
}
\caption{\small
Input parameters used for the numerical analysis.
}
  \label{tab:num-input}
\end{table}

\subsection{Results}

The results obtained for each individual channel as well as for the
total decay width are collected in Tab.~\ref{tab:num-res} and can be
compared to the results obtain previously by BB shown in the first
column. For the $\bar b$-decays the values of all three schemes are
significantly smaller than those obtained by BB. Since BB use an on-shell
(OS) scheme, with masses declared as having some particular values, a
direct comparison is difficult. For  one thing, the
OS corrections enhance the partial decay widths by up to 21\% and 11\%
for $b$ decays, and 68\% and 57\% for $c$ decays, in $\overline{\text{MS}}$ and meson
schemes, respectively.  For instance, for the decay
$\bar b\to \bar cu(\bar s+\bar d)$ we find for the partial width
$\Gamma_{\bar b\to \bar cu(\bar s+\bar d)}$ in
$\text{ps}^{-1}$:\footnote{In the following the results will be shown
  with a two decimal precision, whereas in Tab.~\ref{tab:num-res} and
  Tab.~\ref{tab:errorbudget} three decimal places were kept for a
  better illustration of numerical round-off in the sums.}
\begin{equation}
    \label{eq:bcus-split}
\begin{aligned}
    &0.20= 0.16+0.05-0.01\,,& &\text{($\overline{\text{MS}}$)}\\
    &0.27= 0.27+0.01-0.01\,,& &\text{(meson, Upsilon)}
\end{aligned}
\end{equation}
where the first number on the right-hand side is the OS value and the
second number is the OS correction. The third number corresponds to
non-perturbative corrections which are about 4--5\% of the partial
decay width, which is consistent with the NRQCD counting
$v^2\sim\alpha_s^2\approx 0.04$.

 We only estimate the effect of a non-vanishing strange quark mass on charm decays, that comprise
 60--70\% of the total width. Table~\ref{tab:num-ms-res} gives the
 partial decay rates for $c$ decays including the effect of a non-zero
 strange quark mass. The partial inclusive width for $c$ decays
 includes also a contribution of the doubly Cabibbo suppressed $c\to
 du\bar s$ channel, which is simply estimated as
 $|V_{cd}V_{us}/V_{cs}V_{ud}|^2\Gamma_{c\to su\bar d}$. The strange
 quark mass effect is expected to be suppressed in $\bar b$-decays relative
 to $c$-decays by a factor of
 $\sim(m_c/m_b)^2\sim0.1$, and in any case its computation requires a calculation of QCD corrections  to the decay $b\to
 c\bar cs$ with $m_s\ne0$ which is not available.\footnote{The $b$
   quark decay width  also requires a calculation of QCD corrections  to the
   decay $b\to c\bar u s$, which
   is, however,  Cabibbo suppressed.}  The total decay width of the
 $B_c$ listed in Tab.~\ref{tab:num-ms-res} includes the contributions
 of $\bar b$-decays and of WA and PI listed
 in Tab.~\ref{tab:num-res}.
As a consistency check we confirm that the semi-leptonic $\bar b$-decay agrees with determination of
$V_{cb}$. The PDG's $B^0$ partial semileptonic width is $0.068\pm0.002\,\text{ps}^{-1}$. The
central values of the perturbative contributions to $b\to ce\nu$ in Tab.~\ref{tab:num-res} are
$0.055\,\text{ps}^{-1}$ and $0.071\,\text{ps}^{-1}$ for the $\overline{\text{MS}}$ scheme, and for the meson and Upsilon
schemes, respectively. To compare with data these must be corrected for non-perturbative
effects. The leading non-perturbative correction is of order $1/m_b^2$ in the HQET expansion, and is
given, as a fraction of the total rate, in terms of the non-perturbative matrix elements in HQET,
$\lambda_{1,2}$ , by $-(0.12\lambda_1+0.28\lambda_2)\,\text{GeV}^{-2}$, in the meson and Upsilon
schemes~\cite{Hoang:1998ng,Hoang:1998hm}. Using $\lambda_1=-0.27\pm0.14\,\text{GeV}^2$ and
$\lambda_2=0.12\,\text{GeV}^2$, this gives $0.071\pm0.001\,\text{ps}^{-1}$ in the meson and Upsilon schemes, where the range indicates only  the uncertainty in
$\lambda_1$, exclusive of uncertainties in the perturbative calculation (discussed below).

Similarly, the PDG's $D^0$  inclusive semileptonic width is  $ 0.158\pm 0.003\,\text{ps}^{-1}$. The
sum of
perturbative contributions to $c\to se\nu$ and $c\to d e \nu$  Tab.~\ref{tab:num-ms-res} are
$0.145\,\text{ps}^{-1}$, $0.146\,\text{ps}^{-1}$ and $0.238\,\text{ps}^{-1}$ for the $\overline{\text{MS}}$,  meson and Upsilon
scheme, respectively. Using $(\lambda_1-9\lambda_2)/2m_c^2$ for the fractional non-perturbative correction, we obtain
$0.108\pm0.005\,\text{ps}^{-1}$, $0.103\pm0.005\,\text{ps}^{-1}$ and $0.169\pm0.009\,\text{ps}^{-1}$
in the three different schemes, where the range indicates only the uncertainty in
$\lambda_1$.

\begin{table}
\centering
\renewcommand{\arraystretch}{1.3}
\begin{tabular}{|c|c|c|c|c|}
\hline
  Mode
& BB \cite{Beneke:1996xe}
& $\overline{\text{MS}}$ & meson & Upsilon
\\
\hline\hline
$\bar b\to \bar c u (\bar s+\bar d)$                   & 0.310   &  0.205  &    \multicolumn{2}{|c|}{ 0.266  }
\\
\hline
  $\bar b\to \bar c c (\bar s+\bar d)$               & 0.137          &   0.093      &  \multicolumn{2}{|c|}{ 0.122  }
\\
\hline
$\bar b\to \bar c e\nu$               & 0.075          &    0.053     &   \multicolumn{2}{|c|}{ 0.066   }
\\
\hline
$\bar b\to \bar c \tau\nu$               & 0.018     &  0.010     &  \multicolumn{2}{|c|}{  0.015  }
\\
\hline\hline
$\sum \bar b\to \bar c $               & 0.615      &   0.414      &    \multicolumn{2}{|c|}{ 0.535    }
\\
\hline\hline
$c \to (s+d) u(\bar d+\bar s)$                   & 0.905   &  0.752  &   0.770   & 1.290
  \\
\hline
$c \to (s+d) e\nu$               & 0.162        &  0.161     &    0.162   & 0.250
\\
\hline\hline
$\sum c \to s $               & 1.229          &  1.075  &  1.095   & 1.790
\\
\hline\hline
WA: $\bar b c\to c(\bar s+\bar d)$              & 0.138        &   0.079   &   0.126   & 0.157
\\
\hline
WA: $\bar b c\to \tau\nu$                  & 0.056     &  0.039    &     0.042   & 0.042
\\
\hline
PI                 & $-0.124$            &    $-0.023$    &  $-0.024$ & $-0.017$
\\
\hline\hline
$\Gamma_{B_c}$             &  1.914       &  1.584  &    1.774      & 2.506
\\
\hline\hline
\end{tabular}

\caption{\small
Results for the partial decay rates in ps$^{-1}$. The results from BB in \cite{Beneke:1996xe} in the second column are
compared to the results in this paper given in the $\overline{\text{MS}}$, meson and Upsilon
schemes. The hadronic WA and PI contributions of our results include the QCD penguin contributions, which were neglected in BB.}
  \label{tab:num-res}
\end{table}

\begin{table}
\centering
\renewcommand{\arraystretch}{1.3}
\begin{tabular}{|c|c|c|c|}
\hline
  Mode    & $\overline{\text{MS}}$ & meson & Upsilon
\\
\hline\hline
  $c \to s u \bar d$                  &  0.632  &   0.646   & 1.095
  \\
\hline
  $c \to s u \bar s$                  &  0.033  &   0.032   & 0.057
  \\
\hline
  $c \to d u \bar d$                  &  0.037  &   0.037   & 0.063
  \\
\hline
  $c \to s e\nu$                   &  0.142     &    0.143   & 0.221
\\
\hline
  $c \to d e\nu$                   &  0.008    &    0.008   & 0.013
\\
\hline\hline
  $\sum c \to s $                      &  1.005  &  1.021   & 1.685
\\
\hline\hline
$\Gamma_{B_c}$             &  1.513  &    1.699      & 2.402
\\
\hline\hline
\end{tabular}

\caption{\small
Results for the partial decay rates in ps$^{-1}$ including the effects of non-vanishing strange
quark mass in all three schemes: $\overline{\text{MS}}$, meson and Upsilon
schemes. The ``Total'' row  gives the full decay width.
}  \label{tab:num-ms-res}
\end{table}

\subsection{Uncertainties}
\label{sec:uncertaities}
There are various sources of uncertainty in the calculated width. Use of quark-hadron duality in
the formulation of the OPE method  introduces an irreducible uncertainty that is, moreover,
difficult  to
quantify.  We have little to say about this other than to remind the reader that for semileptonic
decays the situation is much more favourable since one may compute the width in terms of an OPE for
Euclidean momentum region  (that is, for imaginary
time)~\cite{Chay:1990da}. In effect, in absence of new physics effects, this
calculation can provide a test of the validity of the assumption of
quark-hadron duality. Similarly, in the meson and Upsilon schemes we
have neglected the nonperturbative correction to the pole mass, and
again the calculation may be used to provide a test of this assumption.

\begin{table}
\centering
\renewcommand{\arraystretch}{1.3}
\begin{tabular}{|c|c|c|c|c|}
\hline
  parameter $p$
 & $\Delta p/p$ &  $\overline{\text{MS}}$ & meson & Upsilon
 \\
 \hline
   $\overline{m}_b (\overline{m}_b)$              &   0.2\% &  1.815 &  -- & --
   \\
   \hline
     $\overline{m}_c (\overline{m}_c)$              &   0.3\% &  2.798 &  -- & --
\\
\hline
  $\mu$              &   10\% &  $-0.359$ & $ -0.204$  & $-0.112$
\\
\hline
$T$              &   10\% & $-0.029$ &  $-0.034$   & $-0.057$
\\
\hline
$M_{B^*_c}-M_{B_c}$              & 6\%   & 0.012   &  0.015   & 0.016
\\
\hline
$B_{B_c}$              &   30\% &  $-0.004$  &  0.021  & 0.042
\\
\hline
$B'_{B_c}$              &   30\% & 0.065  &  0.060  & 0.030
\\
\hline
$\lambda_1$              &   50\% &  --  & $-0.011$   & 0.017
\\
\hline
$f_{B_c}$              &  1\%  &  0.122 &  0.164  & 0.147
\\
\hline\hline
$V_{cb}$              &  1\%  & 0.644  & 0.769   & 0.575
\\
\hline\hline
\end{tabular}

\caption{\small
Error budget, varying individually the parameter $p$ in the range $\Delta p$, leading to a change in the total rate of $\Delta \Gamma _{B_c}$. The three last columns show the quantity $\frac{\Delta \Gamma_{B_c}}{\Delta p}\frac{p}{\Gamma_{B_c}}$ in the three different mass schemes.
}
  \label{tab:errorbudget}
\end{table}

\subsubsection{Perturbative expansion and QCD-scale uncertainty}
\label{ssec:mu-dep}
The leading contributions to the width of the $B_c$ are from the
perturbatively calculated $\bar b$ and $c$ quark decays. For example,
for $\bar b$ decays, the perturbative calculation of the partial
widths gives
\begin{align*}
   &\qquad (\overline{\text{MS}})& &\text{(meson/Upsilon)}\\
   \Gamma_{b\to c\bar u (s+d)}:~0.21&=0.16+0.05 &0.27&=0.26+0.01\\
   \Gamma_{b\to c\bar c (s+d)}:~0.10&=0.09+0.01 & 0.13&=0.12+0.01\\
   \Gamma_{b\to ce\nu}:~0.06&=0.05+0.01& 0.07&=0.08-0.01\\
   \Gamma_{b\to c\tau\nu}:~0.01&=0.01+0.00& 0.02&=0.02+0.00
\end{align*}
where the first number on the right of each equality is the LO
calculation and the second the order $\alpha_s(m_b)$ correction. These
corrections are seen to be roughly of the expected magnitude, and a
smaller correction is seen in the meson and Upsilon schemes, as expected
from the general considerations presented in Sec.~\ref{sec:Mscheme}.
The improvement is more dramatic for $c$ decays, where we find (for $m_s\ne0$)
\begin{equation*}
   \begin{aligned}
     &\qquad (\overline{\text{MS}})& &\text{(meson)} & &\text{(Upsilon)}\\
     \Gamma_{c\to (s+d) u (\bar s+\bar d)}:~0.76&=0.43+0.33   &  0.78&=0.52+0.27     & 1.35&=1.15+0.20\\
     \Gamma_{c\to (s+d)e\nu}:~0.16&=0.11+0.04&   0.15&=0.13+0.02     & 0.25&=0.30-0.05
   \end{aligned}
\end{equation*}

\begin{figure}
  \begin{center}
    \includegraphics[width=0.45\textwidth]{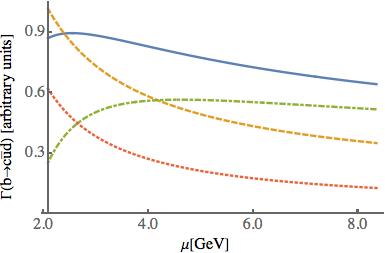}\hspace{0.05\textwidth}
    \includegraphics[width=0.45\textwidth]{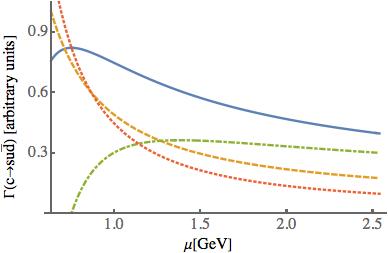}
    \caption{\label{fig:mu-dep} Scale dependence of $\Gamma(b\to cud)$ (left panel) and
      $\Gamma(c\to sud)$ (right panel) in the $\overline{\text{MS}}$ scheme. The solid-blue and dashed-orange lines show the result
      of the LO and NLO
      calculations, respectively. The line in dot-dashed-green shows the LO calculation to which the term with
      the explicit factor of $\alpha_s\ln(\mu)$ in the NLO decay rate is added, displaying cancellation of
      scale dependence to order $\alpha_s$.  The dotted-red line shows the difference between the
      blue and green lines, that is, the NLO decay rate sans the term with an explicit factor of
      $\ln(\mu)$. }
    \end{center}
\end{figure}

The uncertainty from omitting higher order terms in the perturbative expansion is readily estimated
as order $(\alpha_s(\mu_b))^2\sim4\%$ and $(\alpha_s(\mu_c))^2\sim10\%$ for $\bar b$ and $c$ decays,
respectively. This is reflected in the uncertainty introduced by the arbitrary choice of
renormalization scale $\mu$ in the calculations.  For example, the fractional change in the width
for a fractional change in scale, $(\Delta\Gamma/\Gamma)/(\Delta\mu/\mu)$, centered at $\mu=\mu_b$
and $\mu=\mu_c$ for $\bar b$ and $c $ decays, respectively, is $-0.36$, $-0.20$ and $-0.11$ for the
$\overline{\text{MS}}$, meson and Upsilon schemes. It is dominated in all cases by the $\mu$-dependence
of hadronic decays, and among these, of $c$ decays. The scale dependence formally cancels out
to order $\alpha_s$, so the residual dependence is of order $\alpha_s^2$. The variation with respect
to $\ln(\mu)$ of the leading order is of order $\alpha_s$, and this is canceled exactly by the
terms involving $\alpha_s\ln(\mu)$ at NLO. This is shown in Fig.~\ref{fig:mu-dep} that displays the
scale dependence of the NLO calculations of $\Gamma(b\to cud)$ and
$\Gamma(c\to sud)$ in the $\overline{\text{MS}}$ scheme. The dashed (orange) line shows the result of the leading
order calculation, displaying stronger scale dependence than the result of the NLO calculation,
displayed in the solid (blue) line. The latter is the sum of the dashed-dot (green) line  and the
dotted (red) line. This split is intended to demonstrate the approximate cancellation of the $\mu$
dependence that comes in at LO from the Wilson coefficients and through the running mass and NLO
terms with an explicit $\alpha_s\ln(\mu)$: this combination is shown with the dashed-dot
(green) line.  Both green and red
line should be flat up to corrections of order $\alpha_s^2$. We can estimate the uncertainty
due the scale dependence from the range of variation in the figures, about $\pm17\%$ and $\pm35\%$ for
$\bar b$ and $c$ decays, respectively. Since these hadronic decays dominate
the total width of the $B_c$, one can estimate the total uncertainty by weighing these by their
relative importance to the total width, $[(0.202)17\%+(0.743)35\%]/(0.202+0.743)= 31\%.$
This can be compared with the fractional change in scale, $(\Delta\Gamma/\Gamma)/(\Delta\mu/\mu)$,
using $\Delta\mu/\mu\approx\Delta\ln\mu$ which gives $\Delta\Gamma/\Gamma=-0.37\ln2=-0.26$.

Repeating this calculation for the other schemes we infer an uncertainty from order
$\alpha_s^2$/scale dependence of 26\%, 14\% and 8\% in the $\overline{\text{MS}}$, meson and Upsilon schemes.

\subsubsection{Non-relativistic expansion and  Non-perturbative uncertainties}
Additional uncertainties are introduced by the truncation of the non-relativistic expansion. As
already explained we take into account QCD corrections truncated to order $\alpha_s$ while
carrying out the non-relativistic expansion up to relative order $v^4$. Since the power counting in
NRQCD has $v\sim\alpha_s$ this is not fully consistent. However, WA and PI come in first at order $v^6$ and need to be retained because they are numerically
important because the 2-body phase space gives an amplification by a factor of
$16\pi^2$. Corrections to 3-body decays of relative order $v^2$ and $v^4$ are included, so their
numerical effect can be estimated and analyzed. We do not include any QCD corrections to WA and PI,
since they would correspond to higher-order velocity terms which we are neglecting. Furthermore, to have a fully consistent calculation at $\mathcal{O}(v^3)$, three-loop corrections to the rate are required, consisting of three-loop corrections to the matrix elements, a three-loop matching calculation as well as four-loop running. Such corrections are however not available at present and are expected to be smaller than the WA and PI contributions, since they are not enhanced by $16\pi^2$ from the 2-body phase space.

Table~\ref{tab:errorbudget} shows the fractional change in the width per fractional change in
individual parameters. This table can be used to adjust the estimated partial widths in
Tab.~\ref{tab:num-res} by a change in the input parameters in Tab.~\ref{tab:num-input}. The second
column in Tab.~\ref{tab:errorbudget}, ``$\Delta p/p$'', gives our best guess of the fractional
uncertainty in the parameter $p$, so the fractional uncertainties $\Delta\Gamma/\Gamma$ are
estimated by taking the product of the second column with the entries in column 3--5. It is seen
that the uncertainty introduced by the poor knowledge of the matrix elements that go into the
non-relativistic expansion is $\le2\%$ for all entries. The reason is that, in fact, the
non-relativistic corrections arising from non-perturbative matrix elements (that is, other than from
the QCD-perturbative expansion of the free quark decays) are small.

We have already seen in \eqref{eq:bcus-split} that the non-relativistic corrections are small, of
order 4\% for $\Gamma_{\bar b\to \bar c u(\bar s+\bar d)} $. The corrections are slightly larger for
the other $\bar b$ decay modes, up to $\sim10\%$ for the semileptonic modes, but since the width is
dominated by the hadronic modes, the non-relativistic correction to $\Gamma_{\bar b\to\bar c}$ is
4\% and 5\% in the $\overline{\text{MS}}$ and meson/Upsilon schemes, respectively.  Similarly, we find   the non-relativistic correction to $\Gamma_{c\to (s+d)}$ is 6\%, 7\% and
10\% in the $\overline{\text{MS}}$, meson and Upsilon scheme. The small magnitude of the corrections
both suggest the error from truncating the expansion is small, and  the uncertainty introduced from
the limited knowledge of the matrix elements is small, as expected from the previous paragraphs.

The rate of convergence  of the non-relativistic expansion can also be seen by estimating the matrix
element of the  bilinear $\bar QQ$,  that enters at leading order in the OPE, expressed in terms of
NRQCD fields, as given in \eqref{eq:nonRelExp-QQ-2com}. We find, in the Upsilon scheme, the
expansion:
\begin{gather*}
1 - 0.017 - 0.007 + 0.0005 \\
1 - 0.180 + 0.014 +  0.051
  \end{gather*}
where the first line is for $\bar bb$ and the  second for $\bar c c$, and the first, second and
third corrections correspond to the terms $\vec D^2$, $g_s \vec\sigma\cdot\vec B$ and $g_s\vec
D\cdot\vec E$, and  $\vec D^4$, respectively. As expected the non-relativistic expansion converges
faster for $b$ quark decay than for $c$ quarks. We remind the reader
that for the coefficient of the correction terms we have not included
radiative corrections to the coefficient $C_q^{(3)}$ in the OPE; see
the discussion at the end of Sec.~\ref{sec:Cc}. This truncation
represents a correction  $\sim6\%$ in the $\bar c c$ case.

\subsubsection{Parametric and numerical uncertainties}
Additional uncertainties are introduced parametrically, from the uncertainty in the input
parameters. The parametric uncertainty from  $\alpha_s$ is subsumed into the scale uncertainty
discussed in Sec.~\ref{ssec:mu-dep} above. Except for $|V_{cb}|$ and the non-perturbative
parameters that enter the non-relativistic expansion, the input parameters  in
Tab.~\ref{tab:num-input} have negligible uncertainties compared to those that have already been
discussed. The uncertainty from  $|V_{cb}|$ can be read-off Tab.~\ref{tab:errorbudget} that  gives a
fractional uncertainty  in $\Gamma_{B_c}$ slightly smaller than the fractional error in $|V_{cb}|$:
$\Delta\Gamma/\Gamma\lesssim \Delta |V_{cb}|/|V_{cb}|$.

For the $\overline{\text{MS}}$ scheme there are additional parametric
uncertainties from the input value of the quark masses. The resulting
uncertainty is largely due to the fifth power dependence of the
partial widths on the mass of the decaying quark. The conservative
estimate of the
uncertainty assumes that both quark masses deviate in the same direction
from the central value quoted in Tab.~\ref{tab:num-input}. In
 Tab.~\ref{tab:errorbudget} we show the result of varying
$\overline{m}_b(\overline{m}_b)$ and $\overline{m}_c(\overline{m}_c)$
independently, and in the final result we (over)estimate the uncertainty
by adding these linearly, rather than by varying them together.

Additional uncertainties are introduced in the numerical integration of various QCD corrections and
the need to compute at non-zero but sufficiently small gluon mass. We have verified that the error
introduce by the numerical integration and  zero gluon mass extrapolation is much below a percent.

\subsubsection{Strange quark mass}
The non-vanishing of the  strange quark mass reduces the $c$ quark
decay rate, relative to the rate for $m_s=0$, by 7\%, 7\% and 6\% in
the $\overline{\text{MS}}$, meson and Upsilon schemes. A naive estimate of
the effect of the strange mass on the $\bar b$-quark decay is obtained as
this fraction times $(m_c/m_b)^2\sim0.1$ of the $\bar b$-quark width, or
$\Delta\Gamma_b\lesssim0.003\;\text{ps}^{-1}$. More dramatic is the uncertainty derived
from the lesser well determined input,
$\overline{m}_s(2\;\text{GeV})$, listed in
Tab.~\ref{tab:num-input}. Estimating this as $\pm10\%$ of the 6-7\% reduction
in the rate for $c$-quark decay, this results in an uncertainty $\Delta\Gamma_c\sim0.01\;\text{ps}^{-1}$.

\section{Conclusions}
\label{sec:cocl}
The result for the decay rate of the $B_c$ meson, displaying the
contributions by decay channel and for each of the three mass schemes introduced in Sec.~\ref{sec:Mscheme},
 is presented in Tab.~\ref{tab:num-res} for $m_s=0$, and in Tab.~\ref{tab:num-ms-res}  for $m_s\ne0$
 in decay channels involving $c$-quark decay. In Sec.~\ref{sec:uncertaities} we gave a detailed accounting
 of uncertainties in those calculated partial widths. We summarize our results: neglecting light quark masses, including that of the strange quark, our final result for the total
width of the $B_c$   meson in the $\overline{\text{MS}}$-, meson- and
Upsilon-schemes is:
\begin{equation}
\begin{aligned}
  \Gamma^{\overline{\text{MS}}}_{B_c} &= (1.58\pm 0.40|^{\mu}\pm
  0.08|^{\text{n.p.}}\pm 0.02|^{\overline{m}}\pm 0.01|^{V_{cb}})\,\,\text{ps}^{-1}\,, \\
  \Gamma^{\text{meson}}_{B_c} &= ( 1.77\pm 0.25|^{\mu}\pm 0.20|^{\text{n.p.}} \pm 0.01|^{V_{cb}})\,\,\text{ps}^{-1} \,, \\
  \Gamma^{\text{Upsilon}}_{B_c} &= (2.51\pm 0.19|^{\mu}\pm 0.21|^{\text{n.p.}}\pm 0.01|^{V_{cb}})\,\,\text{ps}^{-1} \,,
\end{aligned}
\end{equation}
where we indicate the uncertainties due to scale dependence ($\mu$),
non-perturbative effects (n.p.), the value
of $|V_{cb}|$, and for the $\overline{\text{MS}}$ scheme the input
values of the masses ($\overline{m}$). As discussed above, for $m_s\neq 0$, the central values of the $c$-decay widths get
reduced by about 7\%, leading to the following total decay rates:
\begin{equation}
\begin{aligned}
  \Gamma^{\overline{\text{MS}}}_{B_c} &= (1.51\pm 0.38|^{\mu}\pm 0.08|^{\text{n.p.}}\pm 0.02|^{\overline{m}}  \pm0.01|^{m_s}\pm 0.01|^{V_{cb}})\,\,\text{ps}^{-1}\,, \quad (m_s\neq 0)\\
  \Gamma^{\text{meson}}_{B_c} &= (1.70\pm 0.24|^{\mu}\pm 0.20|^{\text{n.p.}} \pm0.01|^{m_s}\pm 0.01|^{V_{cb}})\,\,\text{ps}^{-1} \,,  \quad (m_s\neq 0)\\
  \Gamma^{\text{Upsilon}}_{B_c} &= (2.40\pm 0.19|^{\mu}\pm 0.21|^{\text{n.p.}} \pm0.01|^{m_s}\pm 0.01|^{V_{cb}})\,\,\text{ps}^{-1} \,, \quad (m_s\neq 0)
\end{aligned}
\end{equation}
For the scale uncertainty we have used Tab.~\ref{tab:errorbudget} with $\Delta\mu/\mu=\ln(2)$. The
 uncertainty due to non-perturbative effects captures the result of the non-relativistic truncation as well as the
uncertainty in the matrix elements. We add linearly the absolute values of the errors from rows 4 --
9 (parameters $T$ through $f_{B_c}$) using for $\Delta p/p$ the values indicated on the second
column, and to this we add linearly the absolute value of the estimate of a non-relativistic
truncation error. The latter is estimated as a fraction, equal to the quark velocity $v_b\approx 0.19 $ or $v_c\approx0.60$,
 of the non-relativistic correction to the $\bar b$ and $c$ decays, respectively, that has been included
 in the partial widths recorded on Tabs.~\ref{tab:num-res} and~\ref{tab:num-ms-res}. Linearly adding
 these uncertainties is the most conservative way to proceed. Adding them instead in quadratures
 leads to the smaller uncertainties $\pm0.05\,\text{ps}^{-1}$,
 $\pm0.11\,\text{ps}^{-1}$ and
 $\pm0.12\,\text{ps}^{-1}$ for both massless and massive strange quarks in the $\overline{\text{MS}}$, meson and Upsilon scheme, respectively.

\begin{table}
\centering
\renewcommand{\arraystretch}{1.3}
\begin{tabular}{|c|c|c|c|}
\hline
 Br(process) & $\overline{\text{MS}}$ &
   meson & {Upsilon} \\
\hline
 $b\to cu(d+s)$ & 13.6 & 15.7 & 11.1 \\
\hline
 $b\to cc(d+s)$ & 6.2 & 7.2 & 5.1 \\
  \hline
 $b\to ce\nu $ & 3.5 & 3.9 & 2.7 \\
  \hline
 $b\to c\tau \nu $ & 0.6 & 0.9 & 0.6 \\
  \hline
 $b \to  c$ & 27.3 & 31.4 & 22.2\\
  \hline
 $c\to su\bar{d}$ & 41.8 & 38.0
         & 45.6 \\
  \hline
 $c\to su\bar{s}$ & 2.1 & 1.9
         & 2.4 \\
  \hline
 $c\to du\bar{d}$ & 2.4 & 2.2 & 2.6 \\
  \hline
 $c\to se\bar{\nu }$ & 9.4 &  8.4 & 9.2 \\
  \hline
 $c\to de\bar{\nu }$ & 0.5 &   0.5 & 0.5 \\
  \hline
 $c \to  s$ & 66.4& 60.1 &    70.2 \\
  \hline
 $bc \to  cs$ & 3.7 & 6.0 &  5.8 \\
  \hline
 $bc \to  \tau \nu $ & 2.6 &  2.5 & 1.8 \\
  \hline
\end{tabular}
\caption{\label{tab:br}Branching fractions in per-cent for $B_c$ semi-inclusive decays in the
 $\overline{\text{MS}}$, meson, and Upsilon schemes. The WA into $cs$ and the PI have been combined
 into a single $bc\to cs$ branching fraction.}
\end{table}

 For either massless or massive strange quarks the three different
 schemes are consistent with each other within their respective
 uncertainties. The difference between the values in the meson and Upsilon scheme can mainly be traced back to the different charm mass used in the charm decays. The wide range spanned by the widths calculated in the three schemes,
 {\it i.e.}, the strong scheme dependence, however calls for an improvement
 of the SM prediction. The main uncertainty, from scale
 dependence, can be reduced by going one order higher in the
 perturbative expansion, {\it i.e.,} using NNLL Wilson coefficients, as well
 as computing the matrix elements up to two loops. While one should see
 convergence of the various schemes on a single value for the
 perturbative partial widths of the underlying $\bar b$ and $c$ decays as
 the calculation is carried out to higher order in $\alpha_s$, we
 expect the Upsilon scheme to converge fastest. Evidence for this was
 presented in Sec.~\ref{sec:schemes-compared} using the known 2-loop
 results for the semileptonic decays of $b$ and $c$ quarks. This is
 also substantiated by the milder scale dependence of the result in the Upsilon
 scheme compared to the other schemes.
 The non-perturbative uncertainty is dominated by our estimate of
 the order $v^3$ corrections. These arise solely from order
 $\alpha_s$ corrections to the Wilson coefficients of order $v^2$
 operators. Additional perturbative corrections to Wilson coefficients
 will have a
 marginal effect on the overall precision. More precise determinations
 of the relevant non-perturbative parameters (for example from lattice
 QCD) might however become important in the future.

It is interesting to note that branching fractions predicted by the
three schemes are in
good agreement with each other, as shown in
Tab.~\ref{tab:br}. This may be interpreted as evidence that the dominant factor
in the differences between
scheme predictions is from the sensitive dependence on masses of the
$\bar b$ and $c$ quarks, which differ vastly among the schemes. We have not
attempted to estimate uncertainties in the branching fractions, which
however are expected to be significantly smaller in the ratios than in
the individual partial widths because of cancellation of correlated
uncertainties.

Previous calculations of the lifetime of $B_c$ mesons yield results
 closer to the experimental measured value than we present
 here. However, the total width depends sensitively on the choice of
 masses for the $b$ and $c$ quarks. An arbitrary choice can be made to
 yield a result close to the experimental value. We have, instead,
 computed systematically, eliminating the pole mass in favour of well
 defined masses and truncating the perturbative expansion
 consistently. This guarantees cancellation of renormalon ambiguities,
that remain present if calculating in terms of ad hoc values for the
 on-shell masses. We find it is interesting to note that the Upsilon
 scheme results have at once the smallest uncertainty and give partial
 widths that are seemingly too large. In particular the semileptonic
 partial width $\Gamma_{c\to (s+d)e\nu}$ is much larger than in both the
 two other schemes and the result of BB. However, as we have remarked
 earlier, it is this higher value that gives excellent agreement with
 the decay width for inclusive semileptonic decays of $D$
 mesons. Similarly, the Upsilon and meson schemes give excellent
 agreement with the inclusive width for semileptonic $B$ meson decay.

 The calculations presented depend on the assumption that
   quark-hadron duality validates the OPE method. Similarly, for the
   Upsilon scheme, and to lesser extend for the meson scheme, the
   nonperturbative contribution to the pole masses have been
   neglected. These assumptions need to be validated through other
   means. Were the predictions to fail the conclusion could well be
   that one or both of these assumptions are not valid.

Finally, the initial motivation for this computation, namely to distinguish NP
 contributions in the $B_c$ life-time remains challenging, and further
 progress on the theory side has to be made before  clear-cut
 conclusions concerning new-physics effects in $\Gamma_{B_c}$ can be
 drawn.

\vspace{1cm}

{\it Acknowledgments} --- {We thank Christine Davies and Aneesh Manohar for useful discussions. J. A. acknowledges
  financial support from the Swiss National Science Foundation (Project No.P400P2\_183838). The work
  of B.G. is
  supported in part by the U.S. Department of Energy Grant No. DE-SC0009919.}

\bibliographystyle{JHEP}
\bibliography{AGBIB}

\providecommand{\href}[2]{#2}\begingroup\raggedright\begin{thebibliography}{10}

\bibitem{Aaij:2014bva}
{\bf LHCb} Collaboration, R.~Aaij et~al., {\it {Measurement of the $B_c^+$
  meson lifetime using $B_c^+ \to J\!/\!\psi \mu^+ \nu_{\mu} X$ decays}},  {\em
  Eur. Phys. J. C} {\bf 74} (2014), no.~5 2839,
  [\href{http://arxiv.org/abs/1401.6932}{{\tt arXiv:1401.6932}}].

\bibitem{Aaij:2014gka}
{\bf LHCb} Collaboration, R.~Aaij et~al., {\it {Measurement of the lifetime of
  the $B_c^+$ meson using the $B_c^+\rightarrow J/\psi\pi^+$ decay mode}},
  {\em Phys. Lett. B} {\bf 742} (2015) 29--37,
  [\href{http://arxiv.org/abs/1411.6899}{{\tt arXiv:1411.6899}}].

\bibitem{Sirunyan:2017nbv}
{\bf CMS} Collaboration, A.~M. Sirunyan et~al., {\it {Measurement of b hadron
  lifetimes in pp collisions at $\sqrt{s} =$ 8 TeV}},  {\em Eur. Phys. J. C}
  {\bf 78} (2018), no.~6 457, [\href{http://arxiv.org/abs/1710.08949}{{\tt
  arXiv:1710.08949}}]. [Erratum: Eur.Phys.J.C 78, 561 (2018)].

\bibitem{Agashe:2014kda}
{\bf Particle Data Group} Collaboration, K.~Olive et~al., {\it {Review of
  Particle Physics}},  {\em Chin. Phys. C} {\bf 38} (2014) 090001.

\bibitem{Bigi:1995fs}
I.~I. Bigi, {\it {Inclusive B(c) decays as a QCD lab}},  {\em Phys. Lett. B}
  {\bf 371} (1996) 105--110, [\href{http://arxiv.org/abs/hep-ph/9510325}{{\tt
  hep-ph/9510325}}].

\bibitem{Beneke:1996xe}
M.~Beneke and G.~Buchalla, {\it {The $B_c$ Meson Lifetime}},  {\em Phys. Rev.
  D} {\bf 53} (1996) 4991--5000,
  [\href{http://arxiv.org/abs/hep-ph/9601249}{{\tt hep-ph/9601249}}].

\bibitem{Chang:2000ac}
C.-H. Chang, S.-L. Chen, T.-F. Feng, and X.-Q. Li, {\it {The Lifetime of $B_c$
  meson and some relevant problems}},  {\em Phys. Rev. D} {\bf 64} (2001)
  014003, [\href{http://arxiv.org/abs/hep-ph/0007162}{{\tt hep-ph/0007162}}].

\bibitem{Kiselev:2000pp}
V.~Kiselev, A.~Kovalsky, and A.~Likhoded, {\it {$B_c$ decays and lifetime in
  QCD sum rules}},  {\em Nucl. Phys. B} {\bf 585} (2000) 353--382,
  [\href{http://arxiv.org/abs/hep-ph/0002127}{{\tt hep-ph/0002127}}].

\bibitem{Gershtein:1994jw}
S.~Gershtein, V.~Kiselev, A.~Likhoded, and A.~Tkabladze, {\it {Physics of B(c)
  mesons}},  {\em Phys. Usp.} {\bf 38} (1995) 1--37,
  [\href{http://arxiv.org/abs/hep-ph/9504319}{{\tt hep-ph/9504319}}].

\bibitem{Gouz:2002kk}
I.~Gouz, V.~Kiselev, A.~Likhoded, V.~Romanovsky, and O.~Yushchenko, {\it
  {Prospects for the $B_c$ studies at LHCb}},  {\em Phys. Atom. Nucl.} {\bf 67}
  (2004) 1559--1570, [\href{http://arxiv.org/abs/hep-ph/0211432}{{\tt
  hep-ph/0211432}}].

\bibitem{Lees:2012xj}
{\bf BaBar} Collaboration, J.~Lees et~al., {\it {Evidence for an excess of
  $\bar{B} \to D^{(*)} \tau^-\bar{\nu}_\tau$ decays}},  {\em Phys. Rev. Lett.}
  {\bf 109} (2012) 101802, [\href{http://arxiv.org/abs/1205.5442}{{\tt
  arXiv:1205.5442}}].

\bibitem{Lees:2013uzd}
{\bf BaBar} Collaboration, J.~Lees et~al., {\it {Measurement of an Excess of
  $\bar{B} \to D^{(*)}\tau^- \bar{\nu}_\tau$ Decays and Implications for
  Charged Higgs Bosons}},  {\em Phys. Rev. D} {\bf 88} (2013), no.~7 072012,
  [\href{http://arxiv.org/abs/1303.0571}{{\tt arXiv:1303.0571}}].

\bibitem{Huschle:2015rga}
{\bf Belle} Collaboration, M.~Huschle et~al., {\it {Measurement of the
  branching ratio of $\bar{B} \to D^{(\ast)} \tau^- \bar{\nu}_\tau$ relative to
  $\bar{B} \to D^{(\ast)} \ell^- \bar{\nu}_\ell$ decays with hadronic tagging
  at Belle}},  {\em Phys. Rev. D} {\bf 92} (2015), no.~7 072014,
  [\href{http://arxiv.org/abs/1507.03233}{{\tt arXiv:1507.03233}}].

\bibitem{Sato:2016svk}
{\bf Belle} Collaboration, Y.~Sato et~al., {\it {Measurement of the branching
  ratio of $\bar{B}^0 \rightarrow D^{*+} \tau^- \bar{\nu}_{\tau}$ relative to
  $\bar{B}^0 \rightarrow D^{*+} \ell^- \bar{\nu}_{\ell}$ decays with a
  semileptonic tagging method}},  {\em Phys. Rev. D} {\bf 94} (2016), no.~7
  072007, [\href{http://arxiv.org/abs/1607.07923}{{\tt arXiv:1607.07923}}].

\bibitem{Aaij:2015yra}
{\bf LHCb} Collaboration, R.~Aaij et~al., {\it {Measurement of the ratio of
  branching fractions $\mathcal{B}(\bar{B}^0 \to
  D^{*+}\tau^{-}\bar{\nu}_{\tau})/\mathcal{B}(\bar{B}^0 \to
  D^{*+}\mu^{-}\bar{\nu}_{\mu})$}},  {\em Phys. Rev. Lett.} {\bf 115} (2015),
  no.~11 111803, [\href{http://arxiv.org/abs/1506.08614}{{\tt
  arXiv:1506.08614}}]. [Erratum: Phys.Rev.Lett. 115, 159901 (2015)].

\bibitem{Hirose:2016wfn}
{\bf Belle} Collaboration, S.~Hirose et~al., {\it {Measurement of the $\tau$
  lepton polarization and $R(D^*)$ in the decay $\bar{B} \to D^* \tau^-
  \bar{\nu}_\tau$}},  {\em Phys. Rev. Lett.} {\bf 118} (2017), no.~21 211801,
  [\href{http://arxiv.org/abs/1612.00529}{{\tt arXiv:1612.00529}}].

\bibitem{Aaij:2017tyk}
{\bf LHCb} Collaboration, R.~Aaij et~al., {\it {Measurement of the ratio of
  branching fractions
  $\mathcal{B}(B_c^+\,\to\,J/\psi\tau^+\nu_\tau)$/$\mathcal{B}(B_c^+\,\to\,J/\psi\mu^+\nu_\mu)$}},
  {\em Phys. Rev. Lett.} {\bf 120} (2018), no.~12 121801,
  [\href{http://arxiv.org/abs/1711.05623}{{\tt arXiv:1711.05623}}].

\bibitem{Li:2016vvp}
X.-Q. Li, Y.-D. Yang, and X.~Zhang, {\it {Revisiting the one leptoquark
  solution to the $R(D^{(*)})$) anomalies and its phenomenological
  implications}},  {\em JHEP} {\bf 08} (2016) 054,
  [\href{http://arxiv.org/abs/1605.09308}{{\tt arXiv:1605.09308}}].

\bibitem{Alonso:2016oyd}
R.~Alonso, B.~Grinstein, and J.~Martin~Camalich, {\it {Lifetime of $B_c^-$
  Constrains Explanations for Anomalies in $B\to D^{(*)}\tau\nu$}},  {\em Phys.
  Rev. Lett.} {\bf 118} (2017), no.~8 081802,
  [\href{http://arxiv.org/abs/1611.06676}{{\tt arXiv:1611.06676}}].

\bibitem{Branco:2011iw}
G.~Branco, P.~Ferreira, L.~Lavoura, M.~Rebelo, M.~Sher, and J.~P. Silva, {\it
  {Theory and phenomenology of two-Higgs-doublet models}},  {\em Phys. Rept.}
  {\bf 516} (2012) 1--102, [\href{http://arxiv.org/abs/1106.0034}{{\tt
  arXiv:1106.0034}}].

\bibitem{Crivellin:2012ye}
A.~Crivellin, C.~Greub, and A.~Kokulu, {\it {Explaining $B\to D\tau\nu$, $B\to
  D^*\tau\nu$ and $B\to \tau\nu$ in a 2HDM of type III}},  {\em Phys. Rev. D}
  {\bf 86} (2012) 054014, [\href{http://arxiv.org/abs/1206.2634}{{\tt
  arXiv:1206.2634}}].

\bibitem{Crivellin:2013wna}
A.~Crivellin, A.~Kokulu, and C.~Greub, {\it {Flavor-phenomenology of
  two-Higgs-doublet models with generic Yukawa structure}},  {\em Phys. Rev. D}
  {\bf 87} (2013), no.~9 094031, [\href{http://arxiv.org/abs/1303.5877}{{\tt
  arXiv:1303.5877}}].

\bibitem{Cline:2015lqp}
J.~M. Cline, {\it {Scalar doublet models confront \ensuremath{\tau} and b
  anomalies}},  {\em Phys. Rev. D} {\bf 93} (2016), no.~7 075017,
  [\href{http://arxiv.org/abs/1512.02210}{{\tt arXiv:1512.02210}}].

\bibitem{Kim:2015zla}
C.~Kim, Y.~W. Yoon, and X.-B. Yuan, {\it {Exploring top quark FCNC within 2HDM
  type III in association with flavor physics}},  {\em JHEP} {\bf 12} (2015)
  038, [\href{http://arxiv.org/abs/1509.00491}{{\tt arXiv:1509.00491}}].

\bibitem{Crivellin:2015hha}
A.~Crivellin, J.~Heeck, and P.~Stoffer, {\it {A perturbed lepton-specific
  two-Higgs-doublet model facing experimental hints for physics beyond the
  Standard Model}},  {\em Phys. Rev. Lett.} {\bf 116} (2016), no.~8 081801,
  [\href{http://arxiv.org/abs/1507.07567}{{\tt arXiv:1507.07567}}].

\bibitem{Wang:2016ggf}
L.~Wang, J.~M. Yang, and Y.~Zhang, {\it {Probing a pseudoscalar at the LHC in
  light of $R(D^{(*)})$ and muon g-2 excesses}},  {\em Nucl. Phys. B} {\bf 924}
  (2017) 47--62, [\href{http://arxiv.org/abs/1610.05681}{{\tt
  arXiv:1610.05681}}].

\bibitem{Hou:1992sy}
W.-S. Hou, {\it {Enhanced charged Higgs boson effects in $B\to\tau \bar\nu ,
  \mu \bar\nu$ and $b\to \tau\bar\nu + X$}},  {\em Phys. Rev. D} {\bf 48}
  (1993) 2342--2344.

\bibitem{Tanaka:1994ay}
M.~Tanaka, {\it {Charged Higgs effects on exclusive semitauonic $B$ decays}},
  {\em Z. Phys. C} {\bf 67} (1995) 321--326,
  [\href{http://arxiv.org/abs/hep-ph/9411405}{{\tt hep-ph/9411405}}].

\bibitem{Kiers:1997zt}
K.~Kiers and A.~Soni, {\it {Improving constraints on $\tan\beta / m_H$ using $B
  \to D \tau\bar\nu$ }},  {\em Phys. Rev. D} {\bf 56} (1997) 5786--5793,
  [\href{http://arxiv.org/abs/hep-ph/9706337}{{\tt hep-ph/9706337}}].

\bibitem{Chen:2006nua}
C.-H. Chen and C.-Q. Geng, {\it {Charged Higgs on $B\to \tau \bar\nu_\tau$ and
  $\bar B\to P(V) \ell \bar\ell$}},  {\em JHEP} {\bf 10} (2006) 053,
  [\href{http://arxiv.org/abs/hep-ph/0608166}{{\tt hep-ph/0608166}}].

\bibitem{Alonso:2015sja}
R.~Alonso, B.~Grinstein, and J.~Martin~Camalich, {\it {Lepton universality
  violation and lepton flavor conservation in $B$-meson decays}},  {\em JHEP}
  {\bf 10} (2015) 184, [\href{http://arxiv.org/abs/1505.05164}{{\tt
  arXiv:1505.05164}}].

\bibitem{Hoang:1998ng}
A.~H. Hoang, Z.~Ligeti, and A.~V. Manohar, {\it {B decay and the Upsilon
  mass}},  {\em Phys. Rev. Lett.} {\bf 82} (1999) 277--280,
  [\href{http://arxiv.org/abs/hep-ph/9809423}{{\tt hep-ph/9809423}}].

\bibitem{Hoang:1998hm}
A.~H. Hoang, Z.~Ligeti, and A.~V. Manohar, {\it {B decays in the upsilon
  expansion}},  {\em Phys. Rev. D} {\bf 59} (1999) 074017,
  [\href{http://arxiv.org/abs/hep-ph/9811239}{{\tt hep-ph/9811239}}].

\bibitem{Bagan:1994zd}
E.~Bagan, P.~Ball, V.~M. Braun, and P.~Gosdzinsky, {\it {Charm quark mass
  dependence of QCD corrections to nonleptonic inclusive $B$ decays}},  {\em
  Nucl. Phys. B} {\bf 432} (1994) 3--38,
  [\href{http://arxiv.org/abs/hep-ph/9408306}{{\tt hep-ph/9408306}}].

\bibitem{Bagan:1994qw}
E.~Bagan, P.~Ball, V.~M. Braun, and P.~Gosdzinsky, {\it {Theoretical update of
  the semileptonic branching ratio of $B$ mesons}},  {\em Phys. Lett. B} {\bf
  342} (1995) 362--368, [\href{http://arxiv.org/abs/hep-ph/9409440}{{\tt
  hep-ph/9409440}}]. [Erratum: Phys.Lett.B 374, 363--364 (1996)].

\bibitem{Krinner:2013cja}
F.~Krinner, A.~Lenz, and T.~Rauh, {\it {The inclusive decay $b \to c\bar{c}s$
  revisited}},  {\em Nucl. Phys. B} {\bf 876} (2013) 31--54,
  [\href{http://arxiv.org/abs/1305.5390}{{\tt arXiv:1305.5390}}].

\bibitem{McNeile:2010ji}
C.~McNeile, C.~Davies, E.~Follana, K.~Hornbostel, and G.~Lepage, {\it
  {High-Precision c and b Masses, and QCD Coupling from Current-Current
  Correlators in Lattice and Continuum QCD}},  {\em Phys. Rev. D} {\bf 82}
  (2010) 034512, [\href{http://arxiv.org/abs/1004.4285}{{\tt
  arXiv:1004.4285}}].

\bibitem{Beneke:1994bc}
M.~Beneke, V.~M. Braun, and V.~I. Zakharov, {\it {Bloch-Nordsieck cancellations
  beyond logarithms in heavy particle decays}},  {\em Phys. Rev. Lett.} {\bf
  73} (1994) 3058--3061, [\href{http://arxiv.org/abs/hep-ph/9405304}{{\tt
  hep-ph/9405304}}].

\bibitem{Luke:1994xd}
M.~E. Luke, A.~V. Manohar, and M.~J. Savage, {\it {Renormalons in effective
  field theories}},  {\em Phys. Rev. D} {\bf 51} (1995) 4924--4933,
  [\href{http://arxiv.org/abs/hep-ph/9407407}{{\tt hep-ph/9407407}}].

\bibitem{Neubert:1994wq}
M.~Neubert and C.~T. Sachrajda, {\it {Cancellation of renormalon ambiguities in
  the heavy quark effective theory}},  {\em Nucl. Phys. B} {\bf 438} (1995)
  235--260, [\href{http://arxiv.org/abs/hep-ph/9407394}{{\tt hep-ph/9407394}}].

\bibitem{Sinkovics:1998mi}
A.~Sinkovics, R.~Akhoury, and V.~I. Zakharov, {\it {Cancellation of $1 / m_Q $
  corrections to the inclusive decay width of a heavy quark}},  {\em Phys. Rev.
  D} {\bf 58} (1998) 114025, [\href{http://arxiv.org/abs/hep-ph/9804401}{{\tt
  hep-ph/9804401}}].

\bibitem{Gray:1990yh}
N.~Gray, D.~J. Broadhurst, W.~Grafe, and K.~Schilcher, {\it {Three Loop
  Relation of Quark (Modified) Ms and Pole Masses}},  {\em Z. Phys. C} {\bf 48}
  (1990) 673--680.

\bibitem{Marquard:2007uj}
P.~Marquard, L.~Mihaila, J.~Piclum, and M.~Steinhauser, {\it {Relation between
  the pole and the minimally subtracted mass in dimensional regularization and
  dimensional reduction to three-loop order}},  {\em Nucl. Phys. B} {\bf 773}
  (2007) 1--18, [\href{http://arxiv.org/abs/hep-ph/0702185}{{\tt
  hep-ph/0702185}}].

\bibitem{Chay:1990da}
J.~Chay, H.~Georgi, and B.~Grinstein, {\it {Lepton energy distributions in
  heavy meson decays from QCD}},  {\em Phys. Lett. B} {\bf 247} (1990)
  399--405.

\bibitem{Pineda:1997hz}
A.~Pineda and F.~Yndurain, {\it {Calculation of quarkonium spectrum and $m_b$,
  $m_c$ to order $\alpha_s^4$}},  {\em Phys. Rev. D} {\bf 58} (1998) 094022,
  [\href{http://arxiv.org/abs/hep-ph/9711287}{{\tt hep-ph/9711287}}].

\bibitem{Melnikov:1998ug}
K.~Melnikov and A.~Yelkhovsky, {\it {The $b$ quark low scale running mass from
  Upsilon sum rules}},  {\em Phys. Rev. D} {\bf 59} (1999) 114009,
  [\href{http://arxiv.org/abs/hep-ph/9805270}{{\tt hep-ph/9805270}}].

\bibitem{Hoang:1998nz}
A.~Hoang, M.~Smith, T.~Stelzer, and S.~Willenbrock, {\it {Quarkonia and the
  pole mass}},  {\em Phys. Rev. D} {\bf 59} (1999) 114014,
  [\href{http://arxiv.org/abs/hep-ph/9804227}{{\tt hep-ph/9804227}}].

\bibitem{Beneke:1998rk}
M.~Beneke, {\it {A Quark mass definition adequate for threshold problems}},
  {\em Phys. Lett. B} {\bf 434} (1998) 115--125,
  [\href{http://arxiv.org/abs/hep-ph/9804241}{{\tt hep-ph/9804241}}].

\bibitem{Manohar:2000dt}
A.~V. Manohar and M.~B. Wise, {\em {Heavy quark physics}}.
\newblock Cambridge University Press, 2000.

\bibitem{Chetyrkin:2000yt}
K.~G. Chetyrkin, J.~H. Kuhn, and M.~Steinhauser, {\it {RunDec: A Mathematica
  package for running and decoupling of the strong coupling and quark masses}},
   {\em Comput. Phys. Commun.} {\bf 133} (2000) 43--65,
  [\href{http://arxiv.org/abs/hep-ph/0004189}{{\tt hep-ph/0004189}}].

\bibitem{Leutwyler:1980tn}
H.~Leutwyler, {\it {How to Use Heavy Quarks to Probe the QCD Vacuum}},  {\em
  Phys. Lett. B} {\bf 98} (1981) 447--450.

\bibitem{Voloshin:1979uv}
M.~Voloshin, {\it {Precoulombic Asymptotics for Energy Levels of Heavy
  Quarkonium}},  {\em Sov. J. Nucl. Phys.} {\bf 36} (1982) 143.

\bibitem{Pak:2008cp}
A.~Pak and A.~Czarnecki, {\it {Heavy-to-heavy quark decays at NNLO}},  {\em
  Phys. Rev. D} {\bf 78} (2008) 114015,
  [\href{http://arxiv.org/abs/0808.3509}{{\tt arXiv:0808.3509}}].

\bibitem{Melnikov:2008qs}
K.~Melnikov, {\it {$O(\alpha_s^2)$ corrections to semileptonic decay $b\to
  c\ell\bar\nu_\ell$}},  {\em Phys. Lett. B} {\bf 666} (2008) 336--339,
  [\href{http://arxiv.org/abs/0803.0951}{{\tt arXiv:0803.0951}}].

\bibitem{Buras:1998raa}
A.~J. Buras, {\it {Weak Hamiltonian, CP violation and rare decays}},  in {\em
  {Les Houches Summer School in Theoretical Physics, Session 68: Probing the
  Standard Model of Particle Interactions}}, pp.~281--539, 6, 1998.
\newblock \href{http://arxiv.org/abs/hep-ph/9806471}{{\tt hep-ph/9806471}}.

\bibitem{Buras:2020xsm}
A.~Buras, {\em {Gauge Theories of Weak Decays}}.
\newblock Cambridge University Press, 6, 2020.

\bibitem{Aebischer:2017gaw}
J.~Aebischer, M.~Fael, C.~Greub, and J.~Virto, {\it {B physics Beyond the
  Standard Model at One Loop: Complete Renormalization Group Evolution below
  the Electroweak Scale}},  {\em JHEP} {\bf 09} (2017) 158,
  [\href{http://arxiv.org/abs/1704.06639}{{\tt arXiv:1704.06639}}].

\bibitem{Lepage:1992tx}
G.~Lepage, L.~Magnea, C.~Nakhleh, U.~Magnea, and K.~Hornbostel, {\it {Improved
  nonrelativistic QCD for heavy quark physics}},  {\em Phys. Rev. D} {\bf 46}
  (1992) 4052--4067, [\href{http://arxiv.org/abs/hep-lat/9205007}{{\tt
  hep-lat/9205007}}].

\bibitem{Grinstein:1997gv}
B.~Grinstein and I.~Z. Rothstein, {\it {Effective field theory and matching in
  nonrelativistic gauge theories}},  {\em Phys. Rev. D} {\bf 57} (1998) 78--82,
  [\href{http://arxiv.org/abs/hep-ph/9703298}{{\tt hep-ph/9703298}}].

\bibitem{Manohar:1997qy}
A.~V. Manohar, {\it {Heavy quark effective theory and nonrelativistic QCD
  Lagrangian to order $\alpha / m^3$}},  {\em Phys. Rev. D} {\bf 56} (1997)
  230--237, [\href{http://arxiv.org/abs/hep-ph/9701294}{{\tt hep-ph/9701294}}].

\bibitem{Dahmen:1981ss}
H.~Dahmen, B.~Scholz, and F.~Steiner, {\it {Infrared Dynamics of Quantum
  Electrodynamics and the Asymptotic Behavior of the Electron Form-factor}},
  {\em Nucl. Phys. B} {\bf 202} (1982) 365--381.

\bibitem{Guberina:1979fe}
B.~Guberina, R.~Peccei, and R.~Ruckl, {\it {Weak Decays of Heavy Quarks}},
  {\em Phys. Lett. B} {\bf 91} (1980) 116--120.

\bibitem{Altarelli:1980fi}
G.~Altarelli, G.~Curci, G.~Martinelli, and S.~Petrarca, {\it {QCD Nonleading
  Corrections to Weak Decays as an Application of Regularization by Dimensional
  Reduction}},  {\em Nucl. Phys. B} {\bf 187} (1981) 461--513.

\bibitem{Altarelli:1991dx}
G.~Altarelli and S.~Petrarca, {\it {Inclusive beauty decays and the spectator
  model}},  {\em Phys. Lett. B} {\bf 261} (1991) 303--310.

\bibitem{Buchalla:1992gc}
G.~Buchalla, {\it {$O(\alpha_s)$ QCD corrections to charm quark decay in
  dimensional regularization with non-anticommuting $\gamma_5$}},  {\em Nucl.
  Phys. B} {\bf 391} (1993) 501--514.

\bibitem{HOKIM1984202}
Q.~Hokim and X.-Y. Pham, {\it Exact one-gluon corrections for inclusive weak
  processes},  {\em Annals of Physics} {\bf 155} (1984), no.~1 202 -- 227.

\bibitem{Cabibbo:1978sw}
N.~Cabibbo and L.~Maiani, {\it {The Lifetime of Charmed Particles}},  {\em
  Phys. Lett. B} {\bf 79} (1978) 109--111.

\bibitem{Nir:1989rm}
Y.~Nir, {\it {The Mass Ratio $m_c / m_b$ in Semileptonic $B$ Decays}},  {\em
  Phys. Lett. B} {\bf 221} (1989) 184--190.

\bibitem{Luke:1992cs}
M.~E. Luke and A.~V. Manohar, {\it {Reparametrization invariance constraints on
  heavy particle effective field theories}},  {\em Phys. Lett. B} {\bf 286}
  (1992) 348--354, [\href{http://arxiv.org/abs/hep-ph/9205228}{{\tt
  hep-ph/9205228}}].

\bibitem{Bagan:1995yf}
E.~Bagan, P.~Ball, B.~Fiol, and P.~Gosdzinsky, {\it {Next-to-leading order
  radiative corrections to the decay $b \to c \bar c s$}},  {\em Phys. Lett. B}
  {\bf 351} (1995) 546--554, [\href{http://arxiv.org/abs/hep-ph/9502338}{{\tt
  hep-ph/9502338}}].

\bibitem{Neubert:1996we}
M.~Neubert and C.~T. Sachrajda, {\it {Spectator effects in inclusive decays of
  beauty hadrons}},  {\em Nucl. Phys. B} {\bf 483} (1997) 339--370,
  [\href{http://arxiv.org/abs/hep-ph/9603202}{{\tt hep-ph/9603202}}].

\bibitem{Kiselev:1994rc}
S.~Gershtein, V.~Kiselev, A.~Likhoded, and A.~Tkabladze, {\it {$B_c$
  spectroscopy}},  {\em Phys. Rev. D} {\bf 51} (1995) 3613--3627,
  [\href{http://arxiv.org/abs/hep-ph/9406339}{{\tt hep-ph/9406339}}].

\bibitem{Bazavov:2018omf}
{\bf Fermilab Lattice, MILC, TUMQCD} Collaboration, A.~Bazavov et~al., {\it
  {Up-, down-, strange-, charm-, and bottom-quark masses from four-flavor
  lattice QCD}},  {\em Phys. Rev. D} {\bf 98} (2018), no.~5 054517,
  [\href{http://arxiv.org/abs/1802.04248}{{\tt arXiv:1802.04248}}].

\bibitem{Colquhoun:2014ica}
B.~Colquhoun, R.~J. Dowdall, C.~T.~H. Davies, K.~Hornbostel, and G.~P. Lepage,
  {\it {$\Upsilon$ and $\Upsilon^{\prime}$ Leptonic Widths, $a_{\mu}^b$ and
  $m_b$ from full lattice QCD}},  {\em Phys. Rev. D} {\bf 91} (2015), no.~7
  074514, [\href{http://arxiv.org/abs/1408.5768}{{\tt arXiv:1408.5768}}].

\bibitem{Lytle:2018evc}
{\bf HPQCD} Collaboration, A.~T. Lytle, C.~T.~H. Davies, D.~Hatton, G.~P.
  Lepage, and C.~Sturm, {\it {Determination of quark masses from
  $\mathbf{n_f=4}$ lattice QCD and the RI-SMOM intermediate scheme}},  {\em
  Phys. Rev. D} {\bf 98} (2018), no.~1 014513,
  [\href{http://arxiv.org/abs/1805.06225}{{\tt arXiv:1805.06225}}].

\bibitem{Tanabashi:2018oca}
{\bf Particle Data Group} Collaboration, M.~Tanabashi et~al., {\it {Review of
  Particle Physics}},  {\em Phys. Rev. D} {\bf 98} (2018), no.~3 030001.

\bibitem{McNeile:2012qf}
C.~McNeile, C.~Davies, E.~Follana, K.~Hornbostel, and G.~Lepage, {\it {Heavy
  meson masses and decay constants from relativistic heavy quarks in full
  lattice QCD}},  {\em Phys. Rev. D} {\bf 86} (2012) 074503,
  [\href{http://arxiv.org/abs/1207.0994}{{\tt arXiv:1207.0994}}].

\bibitem{Dowdall:2012ab}
R.~Dowdall, C.~Davies, T.~Hammant, and R.~Horgan, {\it {Precise heavy-light
  meson masses and hyperfine splittings from lattice QCD including charm quarks
  in the sea}},  {\em Phys. Rev. D} {\bf 86} (2012) 094510,
  [\href{http://arxiv.org/abs/1207.5149}{{\tt arXiv:1207.5149}}].

\end{thebibliography}\endgroup

\end{document}